# Analyzing within Garage Fuel Economy Gaps to Support Vehicle Purchasing Decisions - A Copula-Based Modeling & Forecasting Approach

Behram Wali
Graduate Research Assistant, Department of Civil & Environmental Engineering
The University of Tennessee
bwali@vols.utk.edu

David L. Greene, Ph.D.
Research Professor, Department of Civil and Environmental Engineering &
Senior Fellow, Howard H. Baker, Jr. Center for Public Policy
The University of Tennessee
dgreen32@utk.edu

Asad J. Khattak, Ph.D.
Beaman Professor, Department of Civil & Environmental Engineering
The University of Tennessee
akhattak@utk.edu

Jun Liu, Ph.D.
Travel Demand Modeler
Virginia Department of Transportation (VDOT)
Jun.Liu@VDOT.Virginia.gov

**COMPLETE CITATION:** Wali, B., Greene, D. L., Khattak, A. J., & Liu, J. (2018). Analyzing within garage fuel economy gaps to support vehicle purchasing decisions–A copula-based modeling & forecasting approach. *Transportation Research Part D: Transport and Environment*, *63*, 186-208.

April 26th , 2018

Published in:

Transportation Research Part D

# HIGHLIGHTS

- Unique data reported by customers of the U.S. government website www.fueleconomy.gov are analyzed.

- The study focuses on quantifying the variations of on-road fuel economy relative to official U.S. government ratings (fuel economy difference/gap)

- Proper characterization of the degree of stochastic dependence between the fuel economy difference of pairs of vehicles is sought

- Comprehensive copula based models are developed

- Analysis shows lack of compelling agreement between the fuel economy gaps which could weaken consumers' confidence in making relative comparisons among vehicles.

# Analyzing within Garage Fuel Economy Gaps to Support Vehicle Purchasing Decisions - A Copula-Based Modeling & Forecasting Approach

Behram Wali, David Greene, Asad Khattak, Jun Liu

**Abstract** – A key purpose of the U.S. government fuel economy ratings is to provide precise and unbiased fuel economy estimates to assist consumers in their vehicle purchase decisions. For the official fuel economy ratings to be useful, the numbers must be relatively reliable. This study focuses on quantifying the variations of on-road fuel economy relative to official government ratings (fuel economy gap) and seeks proper characterizations for the degree of stochastic dependence between the fuel economy gaps of pairs of vehicles. By using unique data reported by customers of the U.S. government website www.fueleconomy.gov, the study presents an innovative copula-based joint-modeling and forecasting framework for exploring the complex stochastic dependencies (both nonlinear and non-normal) between the fuel economy gaps of vehicles reported by the same person. While the EPA label estimates are similar to the average numbers reported by website customers, significant, non-linear variation exists in the fuel economy gaps for the two vehicles across the sample. In particular, a positive dependence, characterized by Student-t copula, is observed between the fuel economy gaps of the two vehicles with significant dependencies in the tails of the bivariate distribution; a pair in which one vehicle achieves better (worse) fuel economy is likely to contain a second vehicle getting better (worse) fuel economy as well. However, the results also suggest that the strength of overall association is weak (Kendall's Tau = 0.28). This implies a lack of compelling agreement between fuel economy gaps which could weaken consumers' confidence in making relative comparisons among vehicles.

# 1. INTRODUCTION & BACKGROUND

The Energy Policy and Conservation Act of 1975 requires that the U.S. Department of Energy provides consumers with fuel economy (miles per gallon) ratings for passenger cars and light duty trucks (Rubin et al., 2009; EPA, 2015). The purpose of this requirement is to provide reliable information about vehicle fuel economy ratings to consumers which they can use to compare different vehicles. This information is provided as a fuel economy label on all new passenger cars and light duty trucks sold in the United States and is used in advertisements. The Department of Energy provides two sets of fuel economy estimates, test-cycle and label, for every make, model, engine and transmission configuration (Greene et al., 2017b). The government uses city and highway test-cycle fuel economy numbers to measure compliance with the Corporate Average Fuel Economy (CAFE) and greenhouse gas (GHG) emissions regulations. The label estimates are intended to provide unbiased and reliable fuel economy estimates that assist consumers in their vehicle purchase decisions (EIA, 2016). To account for various driving contexts and travel needs, the label fuel economy estimates are based on three tests (or cycles), which include evaluating higher speeds travel, air conditioning use, and vehicle operation in colder temperatures. The accuracy and unbiasedness of the government's fuel economy label estimates have been a subject of debate for decades. Several studies have documented significant differences between consumer experiences with fuel economy and the government's label fuel economy estimates (McNutt et al., 1978; McNutt et al., 1982; Greene et al., 2006; Lin and Greene, 2011; Greene et al., 2017a). For instance, Greene et al. (2006) found that EPA estimates, while unbiased, are imprecise predictors of individual user's on-road fuel economy with a 95% confidence interval of ±7 miles per gallon (MPG) (Greene et al., 2006). Likewise, a study by Consumer Reports in 2005 concluded a 9% and 18% short fall (difference between on-road fuel economy and EPA label estimates) for conventional gasoline and hybrid vehicles respectively (ConsumerReports,

2013). A 2013 study, however, revealed that the shortfall had decreased to 2% and 10% for gasoline and hybrid vehicles respectively[1] (ConsumerReports, 2013).

Because the EPA fuel economy estimates are the same for all consumers, it is unrealistic to expect that they are providing an exact prediction of each consumer's on-road fuel economy. The government is well-aware of this issue of high variability and includes the following caution on every fuel economy label: "Actual results will vary for many reasons, including driving conditions and how you drive and maintain your vehicle." Nonetheless, as noted in Greene et al. (2006) and Greene et al. (2017), EPA test cycles or label fuel economy values that are imprecise indicators of actual individual on-road fuel economy have important implications (Greene et al., 2006; Greene et al., 2017a). First, a significant portion of drivers will be disappointed with their vehicle's fuel economy performance. Second, drivers may not consider the official EPA ratings as an accurate resource for evaluating the relative fuel economy of similar vehicles (Greene et al., 2006; Greene, 2010). To be useful for relative comparisons, a consumer who experiences a lower fuel economy than the label value on one vehicle should experience a similar result when driving other vehicles. In other words, deviations of real world fuel economy from the label value should be correlated across all vehicles for the same driver.

With these forethoughts in mind, the main objectives of this study are 1) to analyze the variation between reported and label fuel economy numbers for the same individual on different vehicles and 2) to quantify the degree of correlation between fuel economy gaps of vehicles reported by the same individual. The scope of this study is limited to a "My MPG" sample where people entered data on two vehicles in "My MPG" on www.fueleconomy.gov. It is important to note that these are not necessarily two-vehicle households. They are records for two vehicles that are entered by same person. Although it is highly likely that the two-vehicles belong to the same household, we use the term "garage" rather than household to make this distinction. To achieve these objectives, this study performs in-depth statistical analysis of users' self-reported fuel economy data vis-a-vis EPA fuel economy ratings for more than 7000 garages who voluntarily submitted data to the "My MPG" section of the government website[2] www.fueleconomy.gov (Greene et al., 2017a; Greene et al., 2015). Given the research objectives, correctly characterizing the nature of stochastic dependency between fuel economy gaps of the two vehicles within a garage is an important methodological concern. Besides traditional correlation analysis, the study introduces a rigorous copula based methodology in this context which can model complex stochastic dependencies (both nonlinear and non-normal) between the fuel economy gaps under consideration (Genest and Favre, 2007; Trivedi and Zimmer, 2007). By considering a broad spectrum of elliptical, one- and two-parameter Archimedean copulas, rotated/survival variants, and extreme value copulas, we seek a fundamental understanding of the degree of stochastic dependency between fuel economy gaps of two vehicles within a garage. The proposed advanced technique extracts meaningful information embedded in the data (as we will show) which otherwise could not be extracted with simple correlational analysis. In addition to the standard (elliptical and Archimedean) copulas used in this paper, we hope that the introduction of multi-parameter Archimedean copulas and their survival/rotated variants will help translate a diverse suite of copula-devices into the mainstream among the transportation and environment community.

---

[1] Note that this reduction in the gap can be attributed to the change in fuel economy test methods and calculations by the U.S. Environmental Protection Agency in 2006 (https://www.epa.gov/fueleconomy/basic-information-fuel-economy-labeling). The change revised the test methods that U.S. EPA uses to determine fuel economy estimates appearing on window stickers of all passenger vehicles and light-duty trucks.

## 1.1. Conceptual Structure

The current study focuses on exploring and characterizing the degree of stochastic dependency between fuel economy gaps of two vehicles in a particular garage. As discussed above, such an analysis is crucial for understanding the usefulness of official fuel economy label ratings for making relative comparisons among vehicles. While EPA fuel economy ratings are used for other purposes as well, it is assumed in this study that the main purpose of EPA fuel economy label estimates is to provide information to users so that they can make informed comparisons among different vehicles in the marketplace. Figure 1 presents the overall study framework. In this study, fuel economy gap is defined as the ratio of users' self-reported on-road fuel economy and the EPA label ratings (Wali et al., 2018a). As discussed in detail by Wali et al. (2018a), due to the common household, driver, and other unobserved characteristics, several factors that may influence the degree of the gap in fuel economy of one vehicle within a garage may also be common for the second vehicle (Wali et al., 2017b; Wali et al., 2018a). As is evident, this is a bi-dimensional phenomenon and requires the joint modeling of two random variables, i.e., fuel economy gaps of vehicle 1 and vehicle 2 respectively (Wali et al., 2018a).

---

[2] The MyMPG section contains information about real world fuel economy data self-reported by drivers on the MyMPG website operated by the U.S. Department of Energy and U.S. Environmental Protection Agency.

At a basic level, the pairwise dependence between fuel economy gaps of two vehicles within a household can be modelled using classical families of bivariate distributions (Figure 1). As such, the most common models in the classical context can be bivariate normal, log-normal, and/or extreme-value distributions (Boakye et al.; Wali et al., 2018a). However, the classical approach is majorly limited by the fact that the individual behavior of the two variables under consideration (or its transformations) must then be described by the same parametric family of univariate distributions (Genest and Favre, 2007). This also results in strong conceptual implications. For instance, joint normality as of bivariate standard normal distribution may not always exist. Also, the linear form of stochastic dependence which may be implied by bivariate normality may be restrictive (Trivedi and Zimmer, 2007); as it is likely that the nature and degree of stochastic dependence (linear vs. nonlinear) between the fuel economy gaps of the two vehicles may vary across different garages. By varying nature and degree of dependence across the population, we mean that the dependence between factors influencing gaps of vehicle 1 and vehicle 2 may not be linear, i.e., they may possess (as an example) strong central dependence and very weak dependence in distribution tails, or vice versa[3].

---

[3] This will be referred to as the traditional bivariate normality assumption from here onwards.

Therefore, in an attempt to correctly characterize the degree of stochastic dependency between fuel economy gaps of two vehicles within a garage, a rigorous copula based methodology is implemented which can model complex stochastic dependencies (both non linear and non-normal) between the fuel economy gaps under consideration (Genest and Favre, 2007). Copula modeling techniques, which address the aforementioned methodological concerns, have received wide attention in financial markets, actuarial science, and micro-econometrics modeling (Trivedi and Zimmer, 2007), and have recently begun their journey into the travel behavior and traffic safety literature, e.g., (Bhat and Eluru, 2009; Eluru et al., 2010; Yasmin and Eluru, 2013; Ayuso et al., 2016; Wali et al., 2017c).

**(PLACE FIGURE 1 ABOUT HERE)**

## 2. DATA DESCRIPTION

The data for this study are obtained from the joint DOE/EPA website www.fueleconomy.gov, and is the same data used by (Greene et al., 2017a). The government website provides users the opportunity to share their fuel economy estimates for given vehicles (make, model year, engine, and transmission, etc.) (Greene et al., 2017a). Users can provide their fuel economy estimates calculated by different methods; however, the site provides guidance on how to accurately measure the fuel economy. Importantly, the website also provides options for users to compare their fuel economy estimates with the government's official label estimates. Since 2004, a total of 77,126 entries have been made by users. Several consistency checks are run internally to check for errors and plausibility (Goeltz et al., 2005; Lin and Greene, 2011; Greene et al., 2017a). For information about the representativeness, quality, error checking mechanisms, and data elements, interested readers are referred to Greene et al. (2017a) and Wali et al. (2018a) for a detailed discussion.

As the present study focuses on garages for which data is available on two-vehicles, a sub-sample was extracted from the original 77,126 entries. The database includes an anonymous "DriverID" that represents a unique vehicle. Likewise, the variable "GarageID" represents a unique garage, which is likely but not certainly a household. Again, note that the My MPG database does not identify Garage as a unique household. However, as the data related to the two-vehicles are provided by same person, it is likely that the two vehicles are operated within a same household. While the two vehicles likely belong to same household, it is possible that these vehicles are actually driven by different drivers within a household. This may partly contribute to imperfect correlation between the fuel economy gaps of the two vehicles given the (possibly) different driving styles of the household members. However, as several other observed and unobserved factors, such as household income, travel needs, and several environmental and weather related factors, to name a few, would be common to each household, it is reasonable to expect that such common observed and unobserved factors would generate correlation between the fuel economy gaps of the two vehicles.

To identify garages with data on two-vehicles, a query was made to select only those garages that have two distinct driver IDs. As a result, 7244 garages were identified which had data on two-vehicles. Next, we used "DriverID" to identify whether different vehicles belonged to the same garage, and to match the relevant data for each of the two vehicles to their corresponding garages. This resulted in exact matches of the owners' vehicles by make, model, engine, and transmission to their corresponding garages. The data is in a long format where each garage has two rows of data. The first row represents the information on fuel economy, vehicle type, model year, etc. for the first vehicle and second row providing information for vehicle 2.

### 2.1. Outliers in My MPG Data

The My MPG data is characterized by some outliers. Including the outliers in the analysis can distort the modeling results and inferences (Greene et al., 2017a). As such, we conducted a careful outlier analysis to remove observations that can distort the final results. Different methods are used for analyzing the distributions of fuel economy gaps for vehicle 1 and vehicle 2. In particular, observations on fuel economy gaps (for vehicle 1 and 2) were identified that were outside the $\mu \pm 2\ standard\ deviations$ and $\mu \pm 3 \ standard \ deviations$ intervals. Also, we examined the distributions more carefully by constructing Q-Q plots for the trimmed (excluding outliers) and untrimmed (including outliers) samples. The results of outlier analysis are discussed in detail later.

## 2.2. Issue of Self-Reported and Selected Nature of My MPG Sample

The fueleconomy.gov's My MPG database has advantages and disadvantages. The large number of observations and detailed information about the on-road fuel economy with respect to vehicle characteristics make this a unique public source of fuel economy information. The database contains detailed information about users' self-reported fuel economy ratings from all the 50 states in the U.S. The observations for garages with two vehicles are for every model year of vehicles from 1985 to 2014 (discussed later in detail). A weakness of the fueleconomy.gov's My MPG database is the self-selected and self-reported nature of the sample. Given that the fuel economy estimates are self-reported, the data may be a better representation of what people "think" they get rather than "what they actually get." In other words, the issue of self-reporting relates to internal validating – i.e., are we actually measuring what we think we are measuring? Our implicit assumption is that perception measures/relates to reality. However, as mentioned earlier, the fueleconomy.gov website provides valuable guidance to users on how to accurately measure the fuel economy, in addition to several consistency checks, for details see (Greene et al., 2017a). Thus, we believe that the self-reported nature of My MPG sample is unlikely to pose major threats to the study results.

On the other hand, the My MPG sample is self-selected in nature, and which relates to external validity, i.e., can the results obtained from the My MPG sample be generalized to the whole population? As noted in Greene et al. (2017), the distributions of the My MPG data can be examined in contrast to other known data, nonetheless, there exists a possibility of bias in terms of respondents' interest in the topic of fuel economy (Greene et al., 2017a). Using the same database, Greene et al. (2017a) performed tests to detect self-selection bias and found that the parameter estimates of their model of real-world fuel economy were little affected by excluding variables describing driving style or traffic conditions. In addition, self-identified eco-drivers were a small fraction of the participants and the distribution of vehicles in the sample by vehicle type and state corresponded well to the national distributions, although there was some under-representation of light trucks (Greene et al., 2017a). The findings provide evidence that self-selection may not be a serious problem in the My MPG database. For details, see (Greene et al., 2017a).

## 3. STATISTICAL ANALYSIS

The current study focuses on exploring and characterizing the degree of stochastic dependency between fuel economy gaps of two vehicles within a garage. As is evident, this is a bi-dimensional phenomenon and requires the joint modeling of two random variables, i.e., fuel economy gaps of vehicle 1 and vehicle 2 respectively. Given the limitations of classical joint modeling approach (discussed earlier), we first introduce the copula approach and the characteristics and strengths of the different copula dependency devices considered in this study. Next, we discuss in detail different graphical and analytical tools for bivariate exploratory data analysis. The section will be followed by a discussion on goodness-of-fit diagnostics to evaluate the competing nested or non-nested copula based joint models. Finally, we explain the methodology for generating predictions from copula based models and compare the copula-based model predictions with the observed fuel economy gaps of vehicle 1 and 2.

### 3.1. Copula Approach

As stated in an earlier section, modeling and predicting fuel economy gaps of vehicles owned by the same garage/household requires multidimensional (multivariate) modeling of random variables (Wali et al., 2018a). Such analysis has traditionally been formulated using classical multivariate families of distributions which explicitly will assume that the marginals describing the fuel economy gaps for the two vehicles and the multivariate distribution itself belong to the same family. For instance, the assumption of bivariate normality has been used extensively in joint modeling of random variables. In our context, the assumption of bivariate normality will imply that the joint distribution of fuel economy gaps of the two vehicles should be characterized by a bivariate normal distribution (governing the dependence between two fuel economy gaps), whereas the marginals (individual distributions of the two fuel economy gaps) should be approximated by two univariate normal distributions as well. In other words, with the classical multivariate distributions, if the joint density used is bivariate gamma distribution, then the individual behavior of the two variables (fuel economy gaps in our case) must be characterized by a similar parametric family of univariate distributions, and vice versa. In addition, the dependence structure of most traditional multivariate distributions, such as bivariate normal or gamma distributions, explicitly or implicitly assume a linear correlation between the random variables under consideration. To better and more accurately characterize the stochastic dependence between two random variables, it is essential that the modeling of dependence structure is separated out from the modeling of the marginals contributing to the joint density of two random variables. However, doing this is impossible with traditional multivariate distributions, and this along with the restrictive assumption of linear dependence implied by the traditional multivariate distributions, significantly limits our ability to discover the true nature of stochastic dependence between the fuel economy gaps of the two vehicles under consideration.

Copula based joint modeling approach, which was introduced by Sklar (1959), carefully skirts the aforementioned limitations of the traditional multivariate modeling, and allows the analyst to model the dependence structure and univariate margins independently. The copula based joint modeling approach is derived from the Sklar theorem (Sklar, 1959), which states that the joint cumulative distribution function (CDF), $\aleph(x, y)$, for a pair of continuous random variables $(x, y)$ can be formulated as:

$$\aleph(x, y) = C_\theta\{A(x), B(y)\} \tag{1}$$

Where: $A(x)$ and $B(y)$ are the marginal distributions of the fuel economy gaps of vehicle 1 and vehicle 2, $C: [0,1]^2 \to [0,1]$ is the copula device used to tie the $A(x)$ and $B(y)$, and $\theta$ is the dependence parameter governing the stochastic dependency between $x$ and $y$. The $[0,1]$ representation is rooted in the assumption that the pair of continuous random variables has uniform margins in $[0,1]$ i.e., the so-called copula data. Note that this transformation does not affect the inferences made and is solely made to disentangle the marginal distributions from the dependence structure (Sklar, 1959). The copula dependency parameter $\theta$ only depends on the copula device used and not on $A(x)$ and $B(y)$, a property that provides huge flexibility to the modeler (Trivedi and Zimmer, 2007). A valid model for the fuel economy gaps of vehicle 1 and vehicle 2 can arise if $A(x)$, $B(y)$, and $C$ are defined from a parametric family of distributions (Genest and Favre, 2007). If the marginal distributions $A(x)$ and $B(y)$ are normally distributed (choice of margins discussed later), the dependency between the two variables can then be modelled by different families of copulas, $C$. That is, if the two margins are set to be normally distributed based on empirical evidence, then the dependence structure does not need to be characterized by a bivariate normal distribution, and can be modeled using a specific copula governing the true stochastic dependence between the two random variables. Several copulas have been formulated in the literature to model the stochastic dependence between two random variables, with elliptical and Archimedean the two major classes.

### 3.1.1. Elliptical Copulas

Elliptical copulas can be obtained by directly inverting Equation 1, i.e., Sklar's theorem (Sklar, 1959). Given a bivariate distribution function $F$ with invertible margins $A$ and $B$, the elliptical copula family is then represented as:

$$C(x, y) = F\{A^{-1}(x), B^{-1}(y)\} \qquad (2)$$

By choosing from different elliptical distributions, the two most famous bivariate Gaussian copula and bivariate Student-t copula can be obtained (Demarta and McNeil, 2005). Specifically, Gaussian copula has the form:

$$C(x, y) = \Phi_p\{\phi^{-1}(x), \phi^{-1}(y)\} \qquad (3)$$

Where: $p$ is the dependence parameter, $\Phi_p$ is the bivariate standard normal distribution function with correlation $(-1 \leq p \leq 1)$, and $\phi^{-1}$ in Equation 3 is the univariate standard normal distribution function.

Finally, $x$ and $y$ are the two random variables (in our case fuel economy gaps of the two vehicles) in [0,1] interval notation. Likewise, student-t copula has the following notation:

$$C(x,y) = \zeta_{p,v}\{\zeta_v^{-1}(x), \zeta_v^{-1}(y)\} \qquad (4)$$

Compared to the Gaussian copula, the bivariate Student-t distribution is characterized by two parameters, namely the correlation $p$ and the degrees of freedom parameter $v$. And, $\zeta_v^{-1}$ represents the inverse univariate Student-t distribution with $v$ (> 2) degree of freedoms. Following (Nelsen, 2007), the parameter range, Kendall's $\tau$ and tail dependence functions are provided in Table 1.

Gaussian copula is comprehensive copula in that it accounts for both positive and negative dependence between two random variables. The extent of dependence is symmetric around the center. However, the property of asymptotic independence is rooted in Gaussian copula, i.e., regardless of what the actual correlation between two variables may be, extreme events in the tails of the distributions are independent in each of the margin (Nelsen, 2007). In fact, the density function gets very thin in the tails of the distribution with tail dependence approaching 0 (Table 1). In addition, the dependence structure provided by the Gaussian copula is radially symmetric around the center point. This property of Gaussian copula implies that the level of dependence between two variables (fuel economy gaps in our case) is equal in the upper and lower tails of the distribution, and which may not be the case. Finally, with normal marginals assumed, Gaussian copula essentially is a standard bivariate normal distribution. For details, see (Nelsen, 2007; Wali et al., 2017c).

On the other hand, the student-t copula provides a desirable advantage in the sense that it can model the dependence in the tails of the distribution without compromising the flexibility to model stochastic dependence in the center of distribution (Kole et al., 2007). In other words, student-t copula allows more heterogeneity in the joint modeling of fuel economy gaps of the two vehicles under consideration. To quantify the degree of dependence in the tails of the bivariate distribution of $x$ and $y$, the lower and upper tail dependence coefficients ($\lambda_l$ and $\lambda_u$) are used (shown in Table 1). For elliptical copulas, like the student-t copula, the two tail dependence measures ($\lambda_l$ and $\lambda_u$) coincide and can be simply denoted by $\lambda$ (Demarta and McNeil, 2005). The important case occurs when the two tail dependence measures are strictly greater than zero as this will suggest a tendency of the student-t copula to generate joint extreme events, given the fuel economy data at hand. That is, a $\lambda > 0$ will suggest that there is strong upper and lower tail dependencies in the bivariate fuel economy data (Demarta and McNeil, 2005). This property turns out to be very relevant to the problem at hand, i.e., one may expect that the garages who are getting better fuel economy on one vehicle may also get better fuel economy on the second vehicle (and vice versa), i.e., extreme tail events may tend to appear together.

*3.1.2. One-Parameter Archimedean Copulas*

Archimedean family is another popular suite of copulas that have received wide attention in the empirical literature (Genest and Favre, 2007; Nelsen, 2007; Trivedi and Zimmer, 2007). The attraction of this family of copula lies in its closed-form expressions that covers a broad spectrum of stochastic dependency structures (Genest and Rivest, 1993), either symmetrical or asymmetrical (McNeil and Nešlehová, 2009), with the flexibility to assume asymptotic dependence or independence in the tails of the distribution (Charpentier and Segers, 2007). Though the literature on the use of copulas in transportation analysis is small, Archimedean copulas have been mainly used in our field for examining mutlivariate dependencies, e.g., see the seminal work by (Bhat and Eluru, 2009; Bhat et al., 2010) followed by other recent studies (Eluru et al., 2010; Yasmin and Eluru, 2013; Wali et al., 2017c). For details about the properties of Archimedean copulas, interested readers are referred to (Bhat and Eluru, 2009; Wali et al., 2017c). The bivariate Archimedean family of copulas can be formulated as:

$$C(x, y) = \varphi^{[-1]}(\varphi(x) + \varphi(y)) \qquad (5)$$

Where: $\varphi$ is the convex decreasing generator function, continuous in nature (Nelsen, 2007), from $\varphi: [0,1]$ to $\varphi: [0, \infty]$. Additionally, the generator function rests on $\varphi(1) = 0$ with $\varphi^{[-1]}$ in Equation 5 being the pseudo-inverse such that:

$$\varphi^{[-1]}(t) = \begin{cases} \varphi^{[-1]}(t), & 0 \leq t \leq \varphi(0) \\ 0 & \varphi(0) \leq t \leq \infty \end{cases} \quad (6)$$

The generator functions, the range of estimable parameters, expressions for Kendall's $\tau$, and tail dependence formulations for Archimedean copulas are provided in Table 1. Among the Archimedean copulas presented in Table 1, the Frank copula is termed as "comprehensive" copula in the sense that it can account for both positive and negative dependence (if any) between the fuel economy gaps of vehicle 1 and vehicle 2 within the same garage. However, compared to Gaussian copula, the unique feature of Frank copula lies in its ability to capture stronger dependence in the middle of the distribution and weaker dependence in the tails. On the other hand, Clayton, Gumbel, and Joe copulas can only capture positive dependence (albeit with asymmetry in distribution tails), thus called "non-comprehensive copulas". For instance, Joe and Gumbel copulas characterize strong dependence in right tail and weaker dependence in left tail, however, the right tail dependence in Gumbel copula is weaker than in Joe copula. Contrarily, the Clayton copula is ideally suited for stronger dependence in left tail and weak dependence in right tail of the distribution, i.e., it may be useful in cases where there is strong dependence among the two vehicles for which fuel economy is significantly smaller (left tail) than what EPA predicts, however, weaker dependence among the two vehicles for which fuel economy is significantly higher (right tail) than what EPA suggests. For further details about the type of dependency coverage these copulas provide, see (Bhat and Eluru, 2009; Wali et al., 2017c). Following (Coles et al., 1999; Gebremichael and Krajewski, 2007), two versions of asymmetric extensions of Gumbel copula with three parameters, called Tawn type 1 and type 2 copulas, are also tested. In each of the two types, one of the asymmetry parameter is fixed to one, so that the resulting copula destiny is either right (Tawn type 2) or left skewed (Tawn type 1) (Schepsmeier et al., 2017).

*3.1.3. Two-Parameter Hybrid Archimedean Copulas*

The two-parameter bivariate copulas introduced in this paper have attractive characteristics for modeling asymmetric right and left tail dependencies. As noticed above, Archimedean copulas provide a wide range of possible dependence structures. However, the one-parameter Archimedean copulas only allow modeling one type of dependence structure, e.g., strong dependence in right tail and weaker dependence in left tails of the bivariate distribution of fuel economy gaps of two vehicles (Joe copula). This may not be the case, however. The two-parameter Archimedean copulas exhibit very attractive features and provide further flexibility by modeling two different types of dependence structures. As an example, considering the fuel economy gaps in a two-vehicle garage, we may posit that both right and left tail dependencies are stronger, and which cannot be modeled by the one-parameter Archimedean copulas. As such, to model the two different types of dependence structures, we may combine different one-parameter Archimedean copulas, in this case Joe and Clayton copula; Joe copula for modeling the strong dependence in right tail and Clayton copula for modeling the strong dependence in left tail (Wali et al., 2017c). In short, different upper and lower tail dependence structures can be modeled with the two-parameter Archimedean copulas. Thus, in this paper, we introduce the following two parameters hybrid Archimedean copulas: Clayton-Gumbel, Joe-Gumbel, Joe-Clayton, and Joe-Frank copulas. Following (Joe, 2014), we will refer to them as BB1, BB6, BB7, and BB8 respectively. Note that the two-parameter hybrid copulas include the individual copulas (Clayton and Gumbel, BBI; Joe and Gumbel, BB6, and so on) as boundary cases. The generator functions and range of parameters are provided in Table 1. For further details, see Joe (2014) (Joe, 2014) (Section 4.17, Pages 190-210).

*3.1.4. Rotated Archimedean Copulas*

Finally, in addition to the families discussed above, we also consider rotated copulas which in turn provides greater flexibility beyond the ones considered above. When rotated by 180 degrees, we get the corresponding survival copulas that model positive dependence, however with the tail dependence structure switched. Note that the density remains the same but now switched by 180 degrees, and this is possible only for asymmetric copulas which in our case are Clayton, Gumbel, and Joe. As an example, consider Figure 2 which visualizes random samples drawn from Clayton (top plot) and Gumbel (bottom plot) copula with copula parameters governed by Kendall's $\tau$ of 0.75 and -0.75[4]. Recall that Clayton copula is ideal for modeling stronger dependence in the left tail of the bivariate distribution and weaker dependence in the right tail of the distribution (see the first plot in top panel of Figure 2). However, if the analyst observes weak dependence in the left tail and strong dependence in the right tail of the bivariate distribution, the standard Clayton copula cannot be used. In such cases, the concept of rotated Archimedean copulas can be of great value. That is, we can rotate the Clayton copula by 180 degrees, with the resulting dependence of the density still being positive but now with a stronger right tail dependence and weaker left tail dependence (see the third plot in top panel of Figure 2). Following similar interpretation, an un-rotated Gumbel copula is ideal for characterizing a stronger dependence in right tail and weaker dependence in the left tail of the bivariate distribution (shown in the first plot of bottom panel in Figure 2). Given specific bivariate data, a Gumbel copula can be rotated by 180 degrees should an analyst wish to model weaker dependence in the right tail and stronger dependence in the left tail of bivariate distribution[5,6] (see the third plot in bottom panel of Figure 2). As such, the distribution functions of copula rotated by 180 degrees become:

---

[4] Note that for given values of Kendall's $\tau$, the parameters of the copula device can be estimated. For instance, the Kendall's $\tau$ and $\theta$ for Clayton copula are related through the equation, $\text{Kendall's } \tau = \frac{\theta}{\theta+2}$ (see mathematical formulations in Table 1).

[5] At this point, we wish to emphasize that the concept of rotated copulas can also lead us to analyze and model "negative" dependencies using "non-comprehensive" copulas such as Clayton, Gumbel, and Joe, that can otherwise model only "positive" dependencies in its standard (un-rotated) forms. For instance, as mentioned earlier, a standard Clayton copula can model stronger dependence in the left tail and weak dependence in the right tail of the bivariate distribution, with an overall positive dependence between the bivariate distribution (see the upward trend in first plot of the top panel in Figure 2). However, for a negatively correlated bivariate data, if the analyst has to model weaker dependence in the left tail and stronger dependence in the right tail, the Clayton copula can be rotated by 90 degrees which would result in a dependence structure shown in second plot of the top panel in Figure 2. Likewise, for a negatively correlated data, if the aim is to capture stronger dependence in the left tail and weaker dependence in the right tail of bivariate distribution, a standard Clayton copula can be rotated by 270 degrees (see the dependence structure in last plot of top panel in Figure 2). The interesting dependence structures offered by rotated Gumbel copula (shown in bottom panel of Figure 2) can be interpreted in a same fashion. Thus, the concept of rotating copulas by 90 and 270 degrees is of high value in capturing negative dependencies that are otherwise impossible to capture with standard (un-rotated) "non-comprehensive" copulas. Given that the overall dependence structure in our case is positive (discussed later in results section), we do not consider the 90 and 270-degree rotated variants of "non-comprehensive" copulas in this paper.

[6] Altogether, the use of standard elliptical and Archimedean copulas used in this paper, and the two-parameter and rotated variants of Archimedean copulas introduced in this paper allows us to capture a broad spectrum of complex dependence structures between the fuel economy gaps of the two vehicles. A correlational analysis under traditional bivariate normal distribution will imply that the fuel economy gaps of the two vehicles are jointly normally distributed,

$$C_{180}(x,y) = x + y - 1 + C(1-x, 1-y) \quad (7)$$

Note that the traditional correlation coefficient $p$ (Pearson's correlation) is a measure of linear dependence between two random variables, say $X$ and $Y$. The simplistic Pearson's product-moment based correlation coefficient depends on the marginal distributions of the two random variables (Nelsen, 2007), and are thus not applicable to copula models for characterizing the stochastic dependence. Therefore, more robust concordance based measures (such as Kendall's $\tau$) are used to better characterize the dependence between two variables in a copula framework (Nelsen, 2007). Kendall's $\tau$ measures the strength of association between a set of observations of joint random variables $X$ and $Y$. The basic idea is that any pair of random variables $(X_i, Y_i)$ and $(X_j, Y_j)$ (for $x \neq j$) are concordant if the ranks for elements in both pairs match, i.e., $X_i > X_j$ and $Y_i > Y_j$ and vice versa. Contrarily, for $X_i > X_j$ and $Y_i < Y_j$ (or $X_i < X_j$ and $Y_i > Y_j$), the two pairs are discordant. The Kendall's $\tau$ coefficient can then be calculated as $\tau = \frac{(frequency\ of\ concordant\ pairs)-(frequency\ of\ discordant\ pairs)}{\frac{n(n-1)}{2}}$. A perfect agreement between the two rankings will result in $\tau = 1$, whereas a perfect disagreement between the two rankings will result in $\tau = -1$. The reduced form formulae for calculating Kendall's $\tau$ for the copulas considered are provided in Table 1. For derivation of generic expression of Kendall's $\tau$ for different copulas, see Nelson (2007) p.157, and the references therein (Nelsen, 2007).

## 3.2. Estimation of Copula-Models

---

and which may not be the case. Often times, the assumption of bivariate normality in multivariate analysis is made for mathematical and computational convenience rather than as a tool representing the true data structure (Nelsen, 2007; Wali et al., 2017c). Also, a simple correlational analysis based on bivariate normality assumption will indicate a "linear" dependence between the fuel economy gaps of vehicle 1 and 2 (Nelsen, 2007), which is a very restrictive assumption. For instance, the bivariate distribution may possess strong central dependency and relatively weak dependence in the tails of the joint distribution. As another example, the dependencies between fuel economy gaps may be stronger in upper and lower tails, a dependence captured by student-t copula, or with strong dependencies in the right tail and weak dependencies in the left tail of the bivariate distribution, a dependence structure that can be captured with a survival Clayton copula rotated by 180 degrees. We further highlight the benefits of copula-based joint modeling in the results section.

The copula devices can be fitted to the data at hand via several techniques: the maximum pseudo-likelihood method, exact maximum likelihood method, and/or moment-based method based on the inversion of non-parametric dependence measure such as Kendall's $\tau$ (Nelsen, 2007). As our margins are already on uniform scale, we use exact maximum likelihood (ML) estimation technique which provides efficient and satisfactory results (Eluru et al., 2010; Czado et al., 2012). Given a density function for a specific copula, we can use maximum likelihood method to estimate the parameters of the copulas (as discussed above and in Table 1). The method proceeds as follows. Suppose we sample $n$ observations from a multivariate distribution, denoted by $\{(X_{i1}, X_{i2}, X_{i3}, \ldots\ldots, X_{iQ})^T : and\ i = 1,2,3,\ldots, n\}$. Furthermore, the multivariate distribution can be characterized by $Q$ margins ($Q = 2$ in our case) with cumulative distribution function (CDF) $F_i$ and probability distribution function (PDF) $f_i$, $i = 1, \ldots, Q$, and a specific copula with density $c$. Given this setting, the estimable parameters are $\beta$ (vector of marginal parameters) and $\alpha$ (vector of copula parameters). In short, the estimable parameter vector becomes, $\beth = (\beta^T, \alpha^T)^T$, and the log likelihood function to be maximized is:

$$l(\theta) \equiv \sum_{i=1}^{N} \log[C_\theta(F_1(x_{i,1}), \ldots\ldots\ldots, F_Q(x_{i,Q}))|\theta] \qquad (8)$$

For the two dependent variables, i.e., the fuel economy gaps of vehicle 1 and 2, the log-likelihood function reduces to:

$$l(\theta) \equiv \sum_{i=1}^{N} \log[c(F_1(x_{i,1}), F_2(x_{i,2}))|\theta] \qquad (9)$$

Where: $l(\theta)$ is the maximized log-likelihood associated with a particular copula device, $c$ is the density for a specific copula, $\theta$ is the dependence parameter governing the stochastic dependency between the two variables (as $Q = 2$ in our case) for a specific copula, and $F_1(x_{i,1})$, and $F_2(x_{i,2})$ denotes the cumulative distribution functions for the two margins associated with fuel economy of vehicle 1 and 2. Given that the copula parameters are bounded (see Table 1), constrained optimization techniques are used (Nelsen, 2007). Finally, the variance matrix associated with $\beth = (\beta^T, \alpha^T)^T$ is obtained by inverting the Fisher information matrix, and which can be also approximated by the negative of Hessian matrix. The confidence intervals are then established based on the mean and variance estimates obtained above (Schepsmeier et al., 2017).

### 3.3. Graphical Tools for Exploratory Analysis

The true copula characterizing the dependence between the fuel economy gaps of vehicle 1 and vehicle 2 is unknown. Therefore, we first conduct a graphical exploratory analysis to gain insights about the dependence patterns. As a starting point, linear-in-form Pearson correlations are analyzed both for untransformed bivariate data and transformed copula data. As copula methodology disentangles the marginal distributions from the dependence structure (Sklar, 1959), this provides modeler the flexibility to select appropriate distributions for the margins as well. Thus, we analyze the marginal distributions of the fuel economy gaps of vehicle 1 and vehicle 2.

### 3.4. Goodness of Fit/Model Selection
*3.4.1. λ-Function Plot*
To graphically evaluate the goodness of fit of the best-fit copula, we use the $\lambda$-function proposed by Genest and Rivest (1993)(Genest and Rivest, 1993). The $\lambda$-function is unique for each copula family and is formulated as:

$$\lambda(v, \boldsymbol{\theta}) = v - K(v, \boldsymbol{\theta}) \qquad (10)$$

Where: $K(v, \boldsymbol{\theta})$ is the Kendall's distribution function for best-fit copula C and is equal to $K(v, \boldsymbol{\theta}) = P(C(U_1, U_2,)|\boldsymbol{\theta}) \leq v$. $\boldsymbol{\theta}$ (and $\delta$ if two-parameter copula) are dependence parameter(s), $v \in [0,1]$, and the distribution of $(U_1, U_2,)$ is governed by copula C. For Archimedean copulas, closed form expression exists for $\lambda(v, \boldsymbol{\theta})$ and is the quotient of the generation function $\varphi$ (Table 1) and its derivative $\varphi'$ as: $\lambda(v, \boldsymbol{\theta}) = \frac{\varphi(v)}{\varphi'(v)}$. Following (Genest and Rivest, 1993), no closed-form expression exists for Gaussian and Student-t copula, and thus we use simulation to generate theoretical $\lambda$-function based on N = 1000. Finally, to evaluate goodness-of-fit, we compare the empirical $\lambda$-function estimated from data at hand (fuel economy gaps of the two vehicles), theoretical $\lambda$-function for the best-fit copula, and empirical $\lambda$- function superimposed on theoretical $\lambda$-function of the best-fit copula to observe the goodness-of-fit. The closer the $\lambda$-function for a copula superimposes the empirical $\lambda$-function of the data, the better the fit of the particular copula would be. Note that the theoretical $\lambda$-function for the best-fit copula also shows the boundaries of the $\lambda$-function for complete independence (Kendall's $\tau = 0$) and comonotonicity (Kendall's $\tau = 1$ or $= 0$).

Finally, for a formal comparison, we score Akaike Information Criteria (AIC) and Schwarz-Bayesian Information Criteria (BIC) for all 21 competing copula models (Wali et al., 2017a). The model with lowest AIC and BIC is finally selected as the best-fit model (Khattak and Wali, 2017; Wali et al., 2018b). If the AIC and/or BIC happens to be negative, then the best-fit model would be the one with the lowest (more negative) AIC. Note that for comparing non-nested copula models, only BIC is valid (Bhat and Eluru, 2009). The equations for copula versions of AIC and BIC can be found in (Khedun et al., 2014). As a robustness check, we also used the degree of freedom corrected asymptotic tests of Vuong (Vuong, 1989) and Clarke (Clarke, 2007). Doing so resulted in the same choice of best-fit copula model as indicated by AIC and BIC. For details regarding non-nested Vuong test and Clarke test, interested readers are referred to (Vuong, 1989; Clarke, 2007).

**(PLACE TABLE 1 ABOUT HERE)**
**(PLACE FIGURE 2 ABOUT HERE)**

### 3.5. Simulation/Prediction

The proposed methodology can be used for simulating or predicting the fuel economy gaps within a garage/household. Specifically, once a best-fit copula device $C$ is selected and fitted to the bivariate fuel economy gaps, random values can be generated from the chosen copula to assess how the best-fit model simulates/predicts the fuel economy gaps under consideration. In order to simulate dependent bivariate data using the best-fit copula, one must specify the copula family (and associated parameters/shape parameters), marginal distributions for each variable, and the rank correlation parameter governing dependence among variables (Nelsen, 2007). In our case, as normal marginal distributions are assumed given the structure of the data (shown later), we use estimates calculated from the empirical bivariate fuel economy data for the parameters of normal marginals (means and variances). For a 2-dimensional case, a general procedure for simulating observations $(x, y)$ from a pair of random variables $(X, Y)$, with marginals $F_1(x)$, and $F_2(y)$, joint distribution $F_{xy}$, and a best-fit copula $C$ proceeds as follows. By using Sklar's Theorem (Equation 1 and the associated text), and the parameters for given copula device, we only need to generate pairs of observations $(x, y)$ in a uniform interval, i.e., [0,1] representation. Next, following Sklar (1959) and Nelsen (2007) (Sklar, 1959; Nelsen, 2007), for the pairs generated by using the best-fit copula, we can use the *Probability Integral Transform* to transform $(x, y)$ into $(X, Y)$, i.e.,:

$$\begin{cases} X = F_1(x)^{[-1]}(x) \\ Y = F_2(y)^{[-1]}(y) \end{cases} \quad (11)$$

Furthermore, in order to generate a pair $(x, y)$, one can consider the conditional distribution of $Y$ given the case $\{X = x\}$ (Nelsen, 2007), i.e.,:

$$c_x(y) = P\{Y \leq y | X = x\} = \frac{d}{dx} C(x, y) \quad (12)$$

Note that $x$ and $y$ are already sampled from a specified copula family (i.e., the best-fit copula), so applying the appropriate inverse transform (for given distributions of marginals) will lead to the dependent pair at original scale, i.e., original scale of bivariate fuel economy data. For further details about algorithms for prediction/simulation using copula models, see (Nelsen, 2007; Johnson, 2013; Khedun et al., 2014). Finally, we predict the fuel economy gaps of the two vehicles using the best-fit copula model, and undertake a detailed comparison of the simulated/predicted fuel economy gaps with the observed fuel economy gaps for the two vehicles under consideration.

## 4. RESULTS
### 4.1. Outlier Analysis
Previous studies have shown that My MPG data is characterized by outlier observations (Greene et al., 2017a). This is shown in the descriptive statistics in Table 2, where the maximum fuel economy gaps for vehicle 1 and vehicle 2 is 4.178 and 4.585, meaning that there are cases where the two vehicles are getting

317% and 358% better fuel economy that what EPA predicts[7] (see Table 2). Likewise, the minimum fuel economy gaps for vehicle 1 and vehicle 2 is 0.143 and 0.178 respectively; suggesting that there are cases where the two vehicles are getting 85.7% and 82.2% less fuel economy that what EPA predicts. These minimum/maximum statistics seem unreasonable, and can be an outgrowth of some error(s) on part of the user (see statistics for the untrimmed sample in Table 2).

As such, before presenting the descriptive statistics and results of copula-based analysis, we discuss the outlier analysis and present the results. At the first level, we conducted outlier analysis based on number of observations that fall outside the $\mu \pm 2SD$ and $\mu \pm 3SD$. The number of observations falling outside the $\pm 2$ SD and $\pm 3SD$ are shown in Figure 3. For vehicle 1 and 2, 263 and 222 observations were found to be outside the $\mu \pm 2SD$ interval, whereas 71 and 57 observations were found to be outside the $\mu \pm 3SD$ interval respectively (Figure 3). The $\mu \pm 3SD$ threshold is used in removing the outlier observations. As a result, a total of 118 observations were deleted based on the $\mu \pm 3SD$ analysis. Narrower threshold i.e., $\mu \pm 2SD$ was not used as it altered the shape of the bivariate distribution, and creating linear boundaries in the bivariate scatter plot of vehicle 1 and vehicle 2 fuel economy gaps. Note that, removing the outliers based on $\mu \pm 3SD$ did not significantly change the resulting distributions (in terms of means and medians) of the trimmed and untrimmed sample, and are in fact very similar (discussed in sections below).

**(PLACE FIGURE 3 ABOUT HERE)**

Next, to more carefully examine the distributions of fuel economy gaps of vehicle 1 and 2 in the untrimmed and trimmed sample (outliers removed based on $\pm 3SD$ threshold), Q-Q plots are constructed and shown in Figure 4. In particular, the quantiles of observed fuel economy gaps (y-axis) are plotted against the quantiles of corresponding normal distributions (x-axis) for vehicle 1 and 2 in the untrimmed sample (Figure 4a and 4b) and trimmed sample (Figure 4c and 4d). Furthermore, grid lines at 5th, 10th, 25th, 50th, 75th, 90th, and 95th percentiles are provided in the Q-Q plots for visual interpretation.

---

[7] Note that the fuel economy gap is calculated by dividing user self-reported MPG on EPA ratings (Figure 1). Thus, a value of 1.0 will indicate that the users are getting "exactly" the same on-road fuel economy as predicted by the EPA label values, i.e., no gap between on-road fuel economy and EPA label estimates.

As can be seen in Figure 4a and 4b (untrimmed samples), for vehicle 1, there are clear outliers beyond the 95th percentile normal distribution threshold of 1.348 and corresponding 95th percentile threshold of observed data equaling 1.320 (Figure 4a). Likewise, there are outlier observations below the 5th percentile of the observed data and corresponding 5th percentile threshold of normal distribution. Similar observations can be made regarding the fuel economy gaps of vehicle 2 in the untrimmed sample (Figure 4b). These findings suggest that the empirical distributions of fuel economy gaps of vehicle 1 and 2 in the untrimmed sample markedly deviate from normality due to outlier observations, and that too in the tails of the distributions.

To compare the distributions of fuel economy gaps in the trimmed and untrimmed sample, Figure 4c and 4d provides Q-Q plots for the trimmed sample. Note that the Q-Q plots shown in Figure 4c and 4d are constructed after removing observations outside the $\mu \pm 3SD$ threshold (a total of only 118 observations are deleted). It can now be observed that the resulting distributions closely align with the normality line, without problematically altering the distribution of the empirical data. Also, note that the 5th, 50th, and 95th percentile values of the trimmed sample (shown in Figure 4c and 4d) do not significantly differ from the corresponding values in the untrimmed sample (shown in Figure 4a and 4b).

**(PLACE FIGURE 4 ABOUT HERE)**

### 4.2. Descriptive Statistics

Table 2 summarizes the descriptive statistics of key variables (user reported My MPG, EPA ratings, and fuel economy gap (%)) used in this study. As mentioned earlier, the fuel economy gaps are calculated by dividing user reported My MPG by EPA label ratings for the two vehicles, which provides an intuitive measure to quantify the gaps between user reported MPG and EPA ratings. In addition to the outlier analysis presented above, the descriptive statistics (in terms of means and standard deviations) of the trimmed sample (including outliers) and untrimmed sample (excluding outliers) are approximately similar

(Table 2). For the untrimmed sample, the user reported fuel economy for vehicle 1 is (on average) 102.2% of what EPA label ratings suggest, compared to 101.7% for the trimmed sample (Table 2). Likewise, for vehicle 2, the user reported My MPG on-average is 99.8% and 100.5% of the EPA label ratings for the trimmed and untrimmed sample respectively. Importantly, keeping in view the mean values, the standard deviations of the fuel economy gap of vehicle 1 and vehicle 2 suggest significant variations across the sampled garages (Table 2). For instance, the fuel economy gap (%) varies between 42.9% (user getting significantly less than what EPA label estimates suggest) and 160.09% (user getting significantly more than what EPA label estimates suggest) for vehicle 1, and between 42.3% and 159.2% for vehicle 2, for trimmed samples respectively (Table 2). We also observe that the fuel economy shortfalls for vehicle 1 and vehicle 2 are at least 15% (users getting significantly less than what EPA suggest) for 1,075 and 1,182 of the sampled households, respectively. As a result, relative to the fuel costs if official label estimates are used in calculation, the fuel costs for the two-vehicles in a garage can increase significantly because of larger deviations of on-road fuel economy from official label estimates and the significant variations across sampled households.

To provide further insights, descriptive statistics are provided for fuel type, transmission type, and model year in Table 2. As expected, majority 94.3% and 89.5% of the first and second vehicle are gasoline vehicles respectively. Note that vehicle 1 is the older vehicle in the garage based on model year. Thus, it can be observed that the percentage of manual transmission is smaller for vehicle 2 (19.6%) compared to percentage of manual transmission for vehicle 1 (26.2%). Overall, 57.4% and 89.9% of vehicle 1 and vehicle 2 are manufactured in 1999 and later. Also, the associations between My MPG data and EPA label fuel economy estimates across different fuel types (diesel, gasoline, and hybrid) do not exhibit evident bias. For a detailed analysis of the bias in the My MPG data, interested readers are referred to (Greene et al., 2017a). Finally, the sample seems to be reasonably geographically representative in the sense that the observations for two-vehicle garages are spread throughout the 50 states in the US. Figure 5 provides the spatial distribution of the frequencies of observations across all the US states.

**(PLACE TABLE 2 ABOUT HERE)**

**(PLACE FIGURE 3 ABOUT HERE)**

### 4.3. Graphical Dependency Analysis

Before discussing the results of copula analysis of gaps of vehicle 1 and vehicle 2, the results of the graphical exploratory analysis are presented and discussed. The relationship between the fuel economy gaps of vehicle 1 and 2 is visualized in Figure 6a. It is observed that the fuel economy gaps of two vehicles within the same garage are mildly positively correlated with a Pearson correlation of 0.40 (Figure 6a). The density of the bivariate correlation is higher in the middle of the distribution (fuel economy gaps between 0.75 and 0.9) with smaller density elsewhere (Figure 6a). This suggests that garages with fuel economy gaps in between 0.75 and 0.9 for the first vehicle are also likely (albeit with seemingly mild correlation) to experience similar fuel economy gaps for the second vehicle. Nonetheless, the Pearson correlation of 0.402 implies a linear form of stochastic dependence, i.e., the correlation between fuel economy gaps for vehicle 1 and 2 is constant and linear in nature across the population, and which may be unrealistic and unduly restrictive. By non-linear correlation, we mean that the joint distribution of fuel economy gaps may possess (as one example) strong central dependence and very weak dependence in distribution tails, or vice versa, a property that the traditional pearson correlation coefficient does not exhibit.

Figure 6c and 6d visualizes the marginal distributions of the fuel economy gap (with normal density curve overlaid) of vehicle 1 and 2 in the same garage. As can be seen, the marginal distributions of vehicle 1 and 2 are approximately normal. Thus, our choice of characterizing the margins by a normal distribution in subsequent copula analysis seems reasonable (Schepsmeier et al., 2017).

To better detect the dependence between the margins of vehicle 1 and vehicle 2, the original data is now transformed to copula scale (interval [0,1]) which provides deeper and valuable insights[8,9]. A thicker density can be observed now in the tails of the distribution (lower left corner and upper right corner) (Figure 6b). This observation cannot be made from the bivariate scatter plot in Figure 6a. Finally, despite the transformation from original to copula scale, the Pearson correlation coefficient remains the same as it should be[10] (Figure 6b). All of these preliminary findings from the graphical tools suggest that the degree of dependence between fuel economy gaps of vehicle 1 and 2 is likely to vary across the population and is not constant as implied by the traditional bivariate normality assumption. In this case, copula-based dependency analysis can provide deeper and fuller insights about the true nature of stochastic dependency between the fuel economy gaps of the two vehicles.

**(PLACE FIGURE 4 ABOUT HERE)**

### 4.4. Copula Dependency Analysis

In this section, the results of the analysis for selection of most appropriate copula device are presented and discussed. Based on our graphical exploratory analysis discussed above, our a priori hypothesis is that the best-fit copula should be the one with characteristics to capture strong tail-dependencies in the fuel economy gaps as noted earlier. A total of 21 elliptical, one- and two parameter Archimedean, extreme value Gumbel, and their rotated/survival variants are fitted to the fuel economy gaps of vehicle 1 and vehicle 2. Such an analysis is important and can provide deeper information about the stochastic dependencies of the fuel economy gaps under consideration.

---

[8] Figure 6b is constructed using the data on fuel economy gaps of vehicle 1 and 2 transformed into a uniform [0,1] interval, i.e., so-called pseudo observations or copula data (Schepsmeier et al., 2017). The process to create pseudo-observations is as follows. Given "$n$" observations $z_i = (z_{i1}, \ldots, z_{id})^T$, and $i \in \{1, \ldots, n\}$ of a random vector **Z,** the pseudo-observations or copula data can then be defined as $u_{ij} = \frac{r_{ij}}{n+1}$ with $i \in \{1, \ldots, n\}$ and $j \in \{1, \ldots, D\}$. Where, $r_{ij}$ is the rank of $z_{ij}$ among all $z_{kj}$, and $k \in \{1, \ldots, n\}$ (Schepsmeier et al., 2017).

[9] The points in Figure 6a are mapped to points in Figure 6b using the rank-based uniform transformation method explained in footnote 7. As mentioned earlier, any copula operation depends on the pseudo-observations (i.e., original data transformed on a uniform [0,1] interval), and not the original data. This transformation does not affect the interpretation and is mainly done to disentangle the dependence structure between fuel economy gaps of the two vehicles from their marginal distributions. For further details, see (Nelsen, 2007).

The results of all copula models are presented in Table 3. For each copula, Table 3 provides the parameter estimate ($\hat{\theta}$ for one-parameter copulas, and $\hat{\theta}$ and $\hat{\delta}$ for the two-parameter copulas), 95% confidence intervals, and corresponding Kendall's $\tau$ values. Note that the traditional correlational coefficients (ranged between [-1,1]) are not applicable to copula models as such measures (e.g., Pearson correlation) depend on the marginal distributions of the two random variables (Nelsen, 2007). Therefore, more robust concordance based measures (such as Kendall's $\tau$) are used to better characterize the dependence between two variables in a copula framework. The Kendall's $\tau$ correlational values are calculated based on closed-form expressions or through simulation (see Table 1). A perfect agreement between the two rankings for two random variables will result in $\tau = 1$, whereas a perfect disagreement between the two rankings will result in $\tau = -1$. To evaluate the goodness of fit of competing copula models, log-likelihood at convergence, AIC, and BIC are provided in Table 3. The copula model with the lowest AIC and BIC values indicate the best-fit model (Nelsen, 2007; Wali et al., 2017c).

Among all the copula models tested, bivariate models based on Student-t, survival BB1, and BB1 copula devices resulted in the lowest AIC and BIC. Among these three, Student-t resulted in the best fit with lowest AIC and BIC values of -1537.38 and -1523.64 respectively. This finding suggests that the stochastic dependence between the fuel economy gaps of vehicle 1 and 2 can be better characterized by a Student-t copula device. This result is intuitive as strong tail dependence was observed in the exploratory data analysis (discussed above), and t-copula is ideal to better capture the phenomenon of extreme dependent values without compromising the flexibility to model stochastic dependence in the center of the distribution. To visualize the performance of relatively best-fit student-t copula, the left panel of Figure 7 visualizes the $\lambda$-function for best fit t-copula (Nelsen, 2007). In particular, Figure 7a shows the empirical $\lambda$-function for the bivariate fuel economy data, Figure 7b showing the theoretical $\lambda$-function for best-fit Student t-copula (with $\hat{\theta}$ of 0.427 and DF of 5.325), and Figure 7c showing the empirical $\lambda$- function superimposed on the theoretical $\lambda$-function of Student-t copula. For comparison purposes, Figure 7d through 7f shows the theoretical $\lambda$-function, empirical $\lambda$-function for worst-fit Joe copula (with $\hat{\theta}$ of 1.436), and superimposed $\lambda$-functions. As can be seen, Student-t copula approximates the empirical data very closely (Figure 7c) as opposed to the theoretical $\lambda$-function of worst-fit Joe copula largely deviating from the empirical $\lambda$-function of the bivariate data (Figure 7f).

From a practical standpoint, the above findings have important implications as they suggest that garages who are getting better fuel economy on one vehicle are likely to get better fuel economy on the second vehicle too, i.e., right tail dependency. Likewise, a garage getting significantly lower fuel economy (user reported My MPG significantly smaller than EPA ratings) on first vehicle is likely to get significantly lower fuel economy (than what EPA suggests) on the second vehicle too, i.e., left tail dependency. This is reflected in the probability density function (PDF) and cumulative density function (CDF) of the t-copula fitted to the bivariate fuel economy gap data shown in Figure 8. In particular, the tail dependence can be observed in tails of the PDF in Figure 8, with lower and upper tail dependence coefficients of 0.1594 each. The coefficients of tail dependence provide asymptotic measures of the dependence in the tails of the bivariate distribution[11] (Figure 8).

This finding also suggests that the stochastic dependence between the fuel economy gaps of two vehicles is nonlinear (as opposed to assumption of linear dependency in Gaussian copula), and in fact varies across different garages. Such deeper insights cannot be obtained from the traditional bivariate normality assumption as it implies a linear form of dependence between the two pairs, in our case fuel economy gaps of vehicle 1 and vehicle 2. This is evident from Table 3 where Gaussian copula is observed to be statistically inferior (AIC and BIC of -1278.32 and -1271.45) to 12 competing copulas. Also, note that the relatively better performance of BB1 copula is intuitive and not surprising. Recall that BB1 copula is a combination of Clayton and Joe copula, where Clayton copula is ideal to capture the strong left tail dependence and Joe copula ideal for capturing right tail dependence. As such, BB1 copula outperformed other copulas (except t-copula) as the bivariate fuel economy data at hand exhibit strong left and right tail dependencies. Nonetheless, the dependence parameters for all 21 copulas are statistically significant (confidence intervals do not entail zero) suggesting that the two fuel economy gaps of the two vehicles are correlated.

---

[11] As mentioned, the low value of Kendall' $\tau$ suggests an overall weak correlation between the fuel economy gaps of the two vehicles. However, compared to the correlation in the middle of the distribution, the tails of fuel economy gaps exhibit relatively stronger correlation as characterized by the best-fit student-t copula (Figure 8). To quantify the bivariate dependencies in more concrete terms, the coefficients of tail dependence corresponding to copula devices are highly relevant (discussed in section 3.1.1.). In particular, the coefficients of tail dependence provide asymptotic measures of the dependence in the tails of the bivariate fuel economy distribution, i.e., lower and upper tail dependence for the best-fit student-t copula. As shown in Figure 8, the lower and upper tail dependence coefficients for the bivariate fuel economy data are 0.1594 each. In our case, the interpretation of tail dependence coefficients is that it quantifies the probability (chance) of the two random variables (fuel economy gaps of the two vehicles) both taking extreme values (De Kort, 2007). For instance, if a garage/household is experiencing excellent (poor) fuel economy compared to EPA label ratings on their first vehicle, then the probability of the same garage/household experiencing excellent (poor) fuel economy on their second vehicle is 0.1594. In other words, the chance of extreme events happening together is only 15.94 percentage points. Altogether, these findings suggest that while the dependencies between the fuel economy gaps of the two vehicles gets stronger in the tails of the distribution, nonetheless, the probability of two extreme events happening together is still low (probability of 0.1594.).

**(PLACE TABLE 3 ABOUT HERE)**

**(PLACE FIGURE 5 ABOUT HERE)**

Keeping in view the objective of this study, the results in Table 3 also imply that the fuel economy differences for the two vehicles are relatively weakly correlated. The larger variations of user reported My MPG around EPA ratings aside (Greene et al., 2006; Greene et al., 2017a), our analysis based on the sample analyzed suggests that the EPA label ratings may be useful but not as useful as we would like them to be. For example, for the t-copula, the dependence parameter $\hat{\theta}$ of 0.427 and degrees of freedom of 5.325 translates to a Kendall's $\tau$ of 0.281, suggesting a weak positive correlation between fuel economy gaps of vehicle 1 and 2 within the same garage. If all pairs of fuel economy gaps within the garage were concordant (i.e., rank of vehicle 1 gap increases together with the rank of vehicle 2 gap), a Kendall's $\tau$ of 1 would be achieved. A Kendall' $\tau$ of greater than 0.5 would suggest a relatively strong agreement between the fuel economy gaps of the two vehicles in single garage (Abdi, 2007). However, for the data under consideration, the correlation is weak, and which suggests that there is not a compelling agreement between the fuel economy gaps of vehicle 1 and 2. At a basic level, this implies that if a garage is experiencing poor fuel economy (or excellent) compared to EPA label ratings on their first vehicle, the chances that the same garage may get a poor fuel economy (or excellent) on their second vehicle are low. That is, the chance that the two extreme events happen together is only 15.94 percentage points (as explained in footnote 11).

**(PLACE FIGURE 6 ABOUT HERE)**

## 4.5. Bivariate Simulation: Fuel Economy Gaps of Vehicle 1 and 2

In this section, following the procedure explained in section on statistical methods, we present the results of predicting fuel economy gaps using the best-fit student-t copula. Figure 9 shows the bivariate distribution of observed fuel economy gaps (indicated by green color) for the two vehicles under consideration. To examine if the predicted/simulated fuel economy gaps using the student-t copula reasonably fits the observed data, the simulated values are overlaid (indicated by red color) on the observed values in Figure 9. Overall, it can be seen that predictions obtained from the best-fit student-t copula reasonably fits the observed data, especially in the tails of the distribution Figure 9. This is an encouraging result in the sense that information only on the response outcomes is used in the copula modeling framework, without using any explanatory factors that may be correlated with the fuel economy gaps. Also, this demonstrates the potential of copula modeling framework in correctly characterizing the nature of stochastic dependence, and further to use that information in predicting fuel economy gaps under consideration.

**(PLACE FIGURE 7 ABOUT HERE)**

To further evaluate the performance of best-fit student-t copula model in predicting fuel economy gaps, Table 4 presents the detailed summary statistics (mean, median, quantiles, minimum, and maximum) of observed and simulated fuel economy gaps, for vehicle 1 and 2 respectively. Furthermore, in order to quantify how the best-fit copula model performs in predicting fuel economy gap for different model year vehicles, Table 4 also presents the comparison of observed vs. predicted fuel economy gaps by six vehicle model year categories (as defined in Table 2). Finally, following relevant literature (Nelsen, 2007), Mean Absolute Deviance and Root Mean Square Error are reported for quantifying the performance of best-fit copula model in predicting fuel economy gap.

Coming to the results in Table 4, it can be seen that the distributions of observed and simulated fuel economy gaps, for vehicle 1 and 2, for the entire dataset (irrespective of model year categories) reasonably match with each other. For instance, the differences between the median observed and median simulated fuel economy gaps for the two vehicles is 0.008 and 0.004 respectively (Table 4). Likewise, the differences between means, quantiles, and minimum/maximum observed and simulated fuel economy gaps for the two vehicles are not substantial (Table 4). For the entire data, the mean absolute deviance between observed and simulated fuel economy gaps is 0.233 and 0.231 for vehicle 1 and 2 respectively, and which is reasonable (Table 4). In terms of prediction performance by vehicle model year categories, the results suggest minor differences between the distributions of observed and predicted fuel economy gaps (see the distributional statistics in Table 4). Overall, the best-fit student-t copula model fits the observed data by vehicle model year categories reasonably well, as indicated by low mean absolute deviance and root mean squared error statistics (Table 4).

**(PLACE TABLE 4 ABOUT HERE)**

Finally, to better understand the bivariate dependencies between the fuel economy gaps of the two vehicles, the results of conditional bivariate distributional analysis are presented next. In particular, a conditional bivariate analysis of the fuel economy gaps for the two vehicles in a household/garage will provide valuable insights related to the chance of vehicle 1 exceeding the official label estimate when vehicle 2 gets worse than the official label estimate, or vice versa. As an example, if a household is getting at least 10% *better* on-road fuel economy than the label value on their first vehicle, what is the chance that the same household would get at least 10% *worse* on-road fuel economy than the label value on their second vehicle? To quantify such dependencies, the fuel economy gaps

$\left(\text{self} - \text{reported MPG}/\text{EPA label ratings}\right)$ for the two vehicles are classified into four categories as follow:

- Category 1: Vehicle getting at least 10% better on-road fuel economy than the label value (i.e., fuel economy gap is greater than or equal to 1.1)
- Category 2: Vehicle getting at most 10% better on-road fuel economy than the label value (i.e., fuel economy gap between 1 and 1.1)
- Category 3: Vehicle getting at most 10% worse on-road fuel economy than the label value (i.e., fuel economy gap between 0.9 and 1)
- Category 4: Vehicle getting at least 10% worse on-road fuel economy than the label value (i.e., fuel economy gap is less than or equal to 0.9)

Table 5 presents the cross-tabular summary for the observed fuel economy data. To compare the conditional distributions provided by the best-fit model, Table 6 presents the cross-tabular summary for predictions obtained from the best-fit student-t copula model. In both Tables, the number at the top of each cell is the frequency count (shown in white cells), the second number is the row percentage—they sum to 100% going across the table (indicated by light grey cells), i.e., it shows that for a given fuel economy gap category of vehicle 1, what is the distribution of different categories of fuel economy gaps for vehicle 2? Finally, the third number in each cell is the column percentage; they sum to 100% going down the table (indicated by dark grey cells). These numbers show that for a given fuel economy gap category of vehicle 2, what is the distribution of different categories of fuel economy gaps for vehicle 1? Going along a specific row (indicated in light grey cells) or a specific column (indicated in dark grey cells), the row and column percentages can be interpreted as chance, or probabilities if divided by 100. It can be seen that the statistics in Table 6 are fairly close to the conditional distributions based on observed data (Table 5). This illustrates the statistical supremacy of copula based modeling framework (student-t copula in this case) not just in extracting richer insights form the bivariate data at hand but also in predicting joint random variables to a significant level of accuracy. For brevity, we only discuss the conditional distributions obtained from the predicted bivariate fuel economy data. However, the statistics in Table 5 for observed data can be interpreted in a similar fashion.

Several key insights can be obtained from the results in Table 6 as:

- Of all the households who experienced at least 10% better fuel economy than label values on their first vehicle, 44.54% of those households also observed at least 10% better fuel economy than label values on their second vehicle (Table 6).

- However, significant contrasts are also observed in the predicted bivariate fuel economy data. For example, for 13.01% of the households, the predicted fuel economy for vehicle 1 was *at least 10% better* than the label values, whereas, the predicted fuel economy for vehicle 2 *was at least 10% worse* than the label values (Table 6). In addition, for 18.88% of the households, the fuel economy for vehicle 1 was at least 10% better than the label values, but the fuel economy for vehicle 2 *was at most 10% worse* than the label values. Altogether, for 31.8% of the households, the fuel economy on vehicle 1 was at least 10% better than the label value while the fuel economy on vehicle 2 was less than the corresponding label values (Table 6).

- Likewise, for 12.9% of the households, the second vehicle experienced *at least 10% better fuel economy* than the label value while the fuel economy for vehicle 1 was *at least 10% worse* than the label values (Table 6). In addition, similar to the conditional distributions based on vehicle 1, for 27.3% of the households, vehicle 2 experienced at least 10% better on-road fuel economy than the label value, whereas, the fuel economy on vehicle 1 for the same households was less than their corresponding label values (Table 6).

- These results suggest that there exists a significant chance that a household experiencing better fuel economy than the label value on one of their vehicles will experience worse fuel economy than the label value on the other vehicle.

# 5. CONCLUSIONS/PRACTICAL IMPLICATIONS

The EPA label fuel economy estimates are important for two main reasons: 1) they provide fuel economy numbers used by the government to enforce CAFE and GHG emissions for light-duty vehicles, and 2) they provide information that helps consumers' car purchase decisions (vehicle choices). The accuracy, unbiasedness, and usefulness of the government's fuel economy estimates have been a subject of debate for decades. While the high variability of on-road fuel economy estimates relative to official fuel economy ratings have been recognized and analyzed in the past, efforts to understand the usefulness of official fuel economy estimates for making comparisons among vehicles through empirical analysis are rare. Having said this, the present study focused on two important questions:

- How much intra-garage variations of fuel economy gaps exist for vehicles owned by the same garage?
- What is the strength and nature of the correlation between the fuel economy gaps of vehicles owned by the same garage?

For the official fuel economy ratings to be useful for making relative comparisons among vehicles, we need a fundamental understanding regarding the degree of stochastic dependence between fuel economy gaps within garages. That is, relative to the official EPA ratings, if the garage experiences a high or low fuel economy on their first vehicle, can they also expect similar results on their second vehicle? In our attempt to answer this question, we conducted an in-depth statistical analysis of users' self-reported fuel economy data vis-a-vis EPA fuel economy ratings for more than 7000 two-vehicle garages who voluntarily submitted data to the "My MPG" section of the government website www.fueleconomy.gov. Correctly characterizing the degree of stochastic dependency between the fuel economy gaps of two-vehicle garages is an important methodological concern. As such, besides the traditional correlational analysis, we introduce rigorous copula based methodology in this context which can model complex stochastic dependencies (both nonlinear and non-normal) between the fuel economy gaps under consideration. By considering a broad spectrum of elliptical, one- and two-parameter

Archimedean copulas (and its survival variants), and extreme value copulas, the proposed technique extracts information embedded in data which otherwise could not be extracted with simple correlational analysis.

The analysis reveals that the EPA label estimates provide a reasonable estimate of the average user-reported on-road fuel economy. However, the study observes significant variation in the fuel economy gaps for the two vehicles across the sampled garages. For instance, the mean fuel economy gaps $\left(user\ reported\ MPG/_{EPA\ label\ estiamtes} * 100\right)$ for vehicle 1 and vehicle 2 are 101.6% and 99.8%, with Interquartile ranges of 21% and 20.4%, for vehicle 1 and 2, respectively. The results from copula analysis suggest that the stochastic dependence between the fuel economy gaps of two vehicles is not linear (as opposed to assumption of linear dependency in Gaussian copula), and in fact varies across different garages. In particular, the study shows dependencies in the tails of the bivariate distribution, i.e., garages who are getting better fuel economy on one vehicle are likely to get better fuel economy on the second vehicle as well. Given the complex nature of the correlation, Student-t copula (among 21 different copula models) is the best-fit for the data at hand and provides meaningful insights that were otherwise not possible from the Gaussian copula or the traditional bivariate normality assumption. Analysis of the deviations for pairs of vehicles in the same garage suggests that the EPA label ratings are useful, but not as useful as one would like them to be. The strength of these associations is weak overall, creating uncertainty even in relative comparisons among vehicles.

We emphasize that the EPA label values are primarily intended to inform consumers before vehicle choice decisions. The EPA label estimates would be of greatest value if the numbers were accurate for every individual and for every vehicle (a tall order though). In that case, consumers could accurately estimate fuel costs for all vehicles (subject to uncertainty about fuel prices). Given that this is not possible for a single label number, it would be desirable for the label value to be proportional to the actual fuel economy a consumer would get, with the proportionality constant for all vehicles (i.e., a constant known or even unknown ratio). For a known ratio, the consumer could still calculate accurate fuel costs with a little extra effort. While the ratio (i.e., the percentage difference between EPA label and on-road user reported MPG) is known in this study, the fuel economy gaps (ratio) vary significantly across the vehicles and garages (as discussed earlier). Given that the differences exist, the degree of usefulness depends on how strongly the differences are correlated. Our analysis shows that the discrepancies (variations in fuel economy gaps) are large and the correlation is weak. For example, the probability of extreme events, both vehicles in household getting excellent on-road fuel economy or both vehicles getting poor on-road fuel economy, happening together is 0.1594. Thus, the label values may even produce incorrect rankings of vehicles for a specific consumer. Furthermore, the conditional distributional analysis presented in this paper showed that the chance of a household getting at least 10% better fuel economy than the label value on their first vehicle while getting lower fuel economy (than the label value) on their second vehicle is 31.8%. These findings, when combined with the others discussed in the paper, suggest that there exists a significant chance that a household experiencing better fuel economy than the label value on one of their vehicles will experience worse fuel economy than the label value on their other vehicle.

As the key focus of the present study was to investigate the stochastic dependence between fuel economy gaps, we employed copula modeling techniques to model unconditional fuel economy gaps, as is typically done in other fields (Gebremichael and Krajewski, 2007; Genest and Favre, 2007; Khedun et al., 2014). In the future, it will be interesting to jointly model the fuel economy gaps conditional on different garage specific explanatory factors, i.e., by incorporating the effects of covariates (such as the effect of make and model year) into the margins and/or the copula parameters. Another natural extension of the current work would be to examine the stochastic dependence among fuel economy gaps for garages with more than two vehicles.

## 6. ACKNOWLEDGEMENT

This paper is based upon work supported by the U.S. Department of Energy and Oak Ridge National Laboratory. The support of University of Tennessee's Transportation Engineering and Science Program and Initiative for Sustainable Mobility, a campus-wide organized research unit, is gratefully appreciated. Any opinions or recommendations expressed in this paper are those of the authors. The authors would also like to recognize the contributions of Ms. Alexandra Boggs, Ms. Megan Lamon, and Mr. Jerome Zachary in proof-reading earlier versions of the manuscript.

# LIST OF FIGURES

# LIST OF TABLES

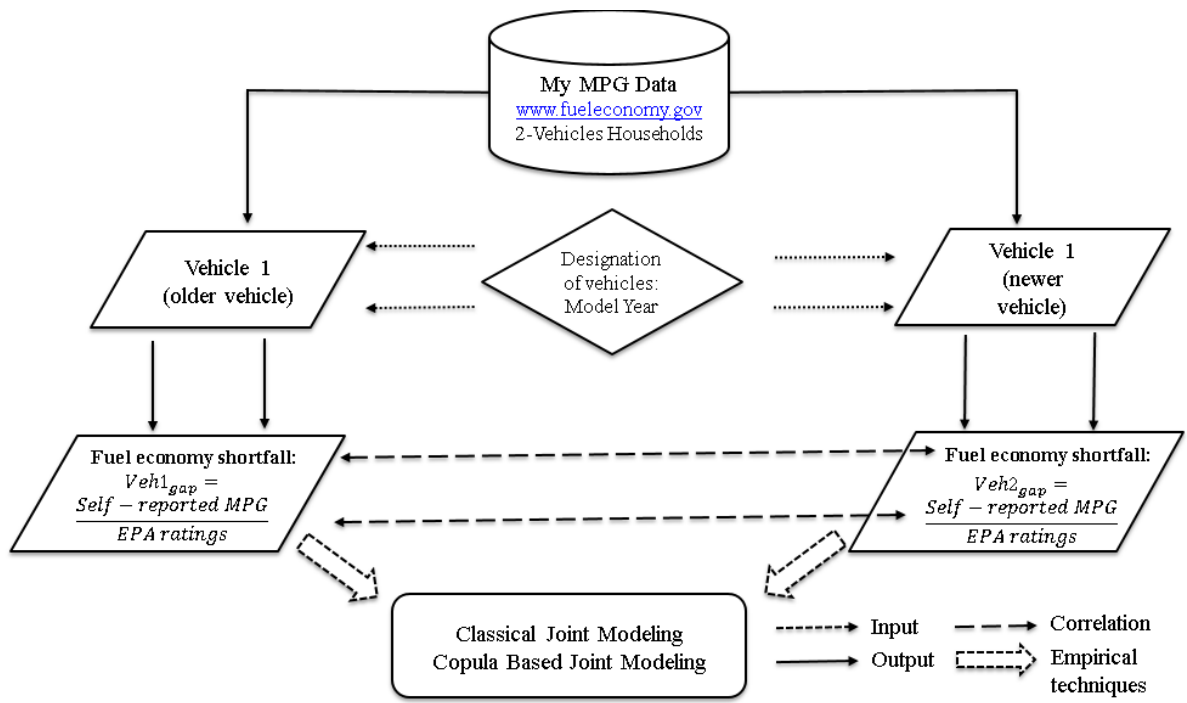

**FIGURE 1 Study framework**

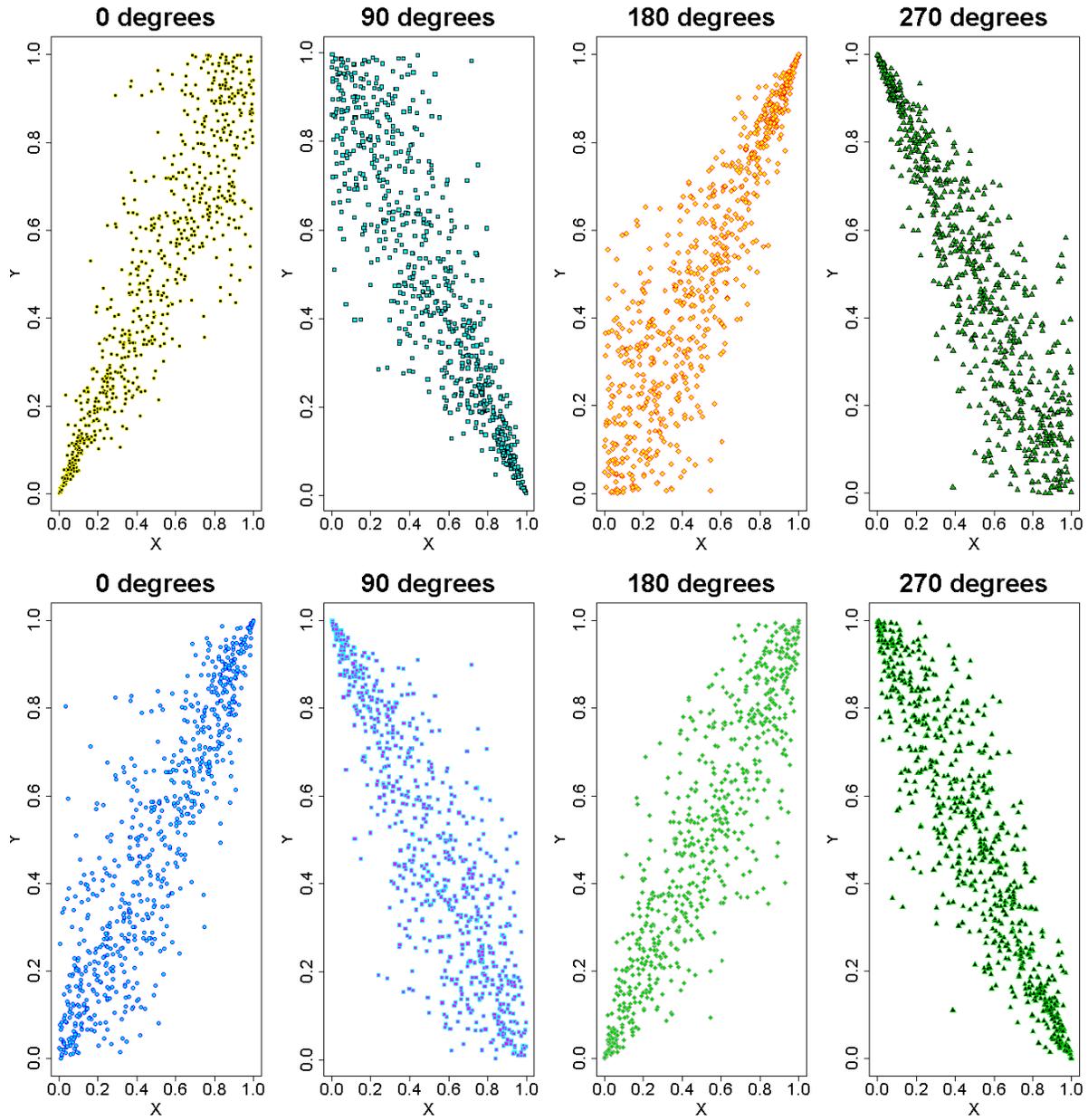

**FIGURE 2: Rotated Archimedean Copulas**

Note: 700 samples simulated from Clayton (top plot) and Gumbel (bottom plot) copulas rotated by 0, 90, 180, and 270 degrees with copula parameters corresponding to Kendall's $\tau$ of 0.75 and -0.75 for positive (0 and 180-degree rotation) and negative dependence (90 and 270-degree rotation) respectively.

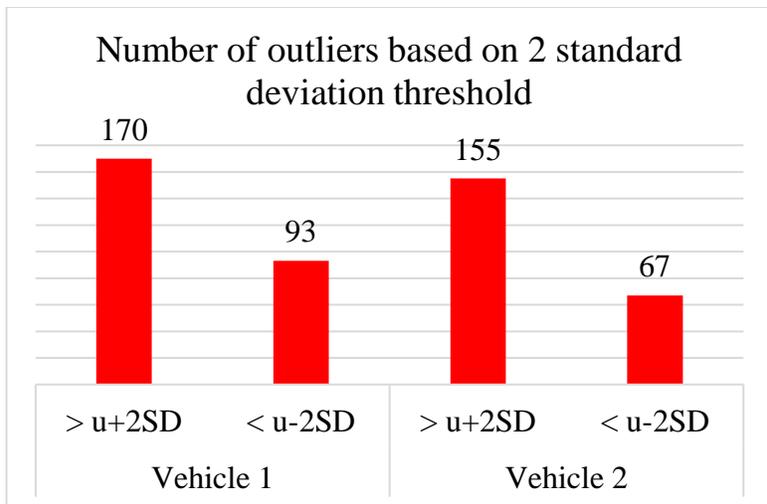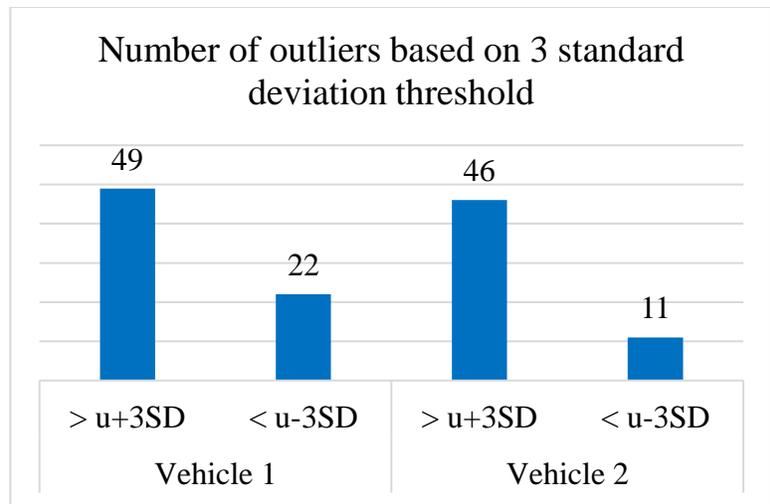

**FIGURE 3:** Number of Outlier Observations For Fuel Economy Gaps Falling Outside The 2 Standard Deviations And 3 Standard Deviations

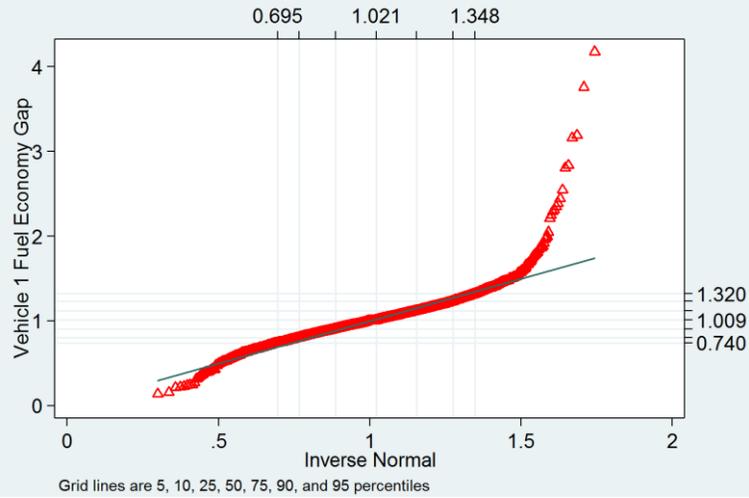
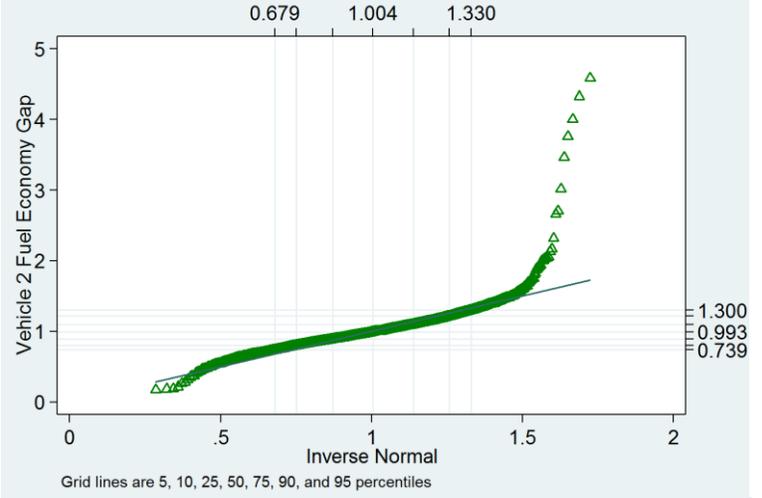
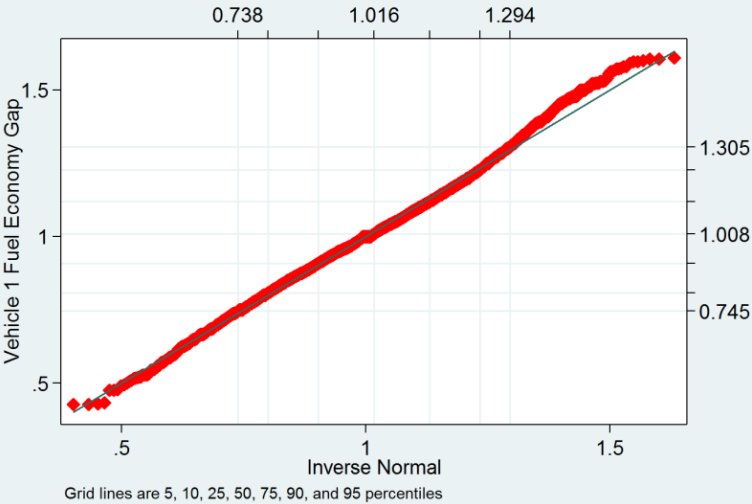
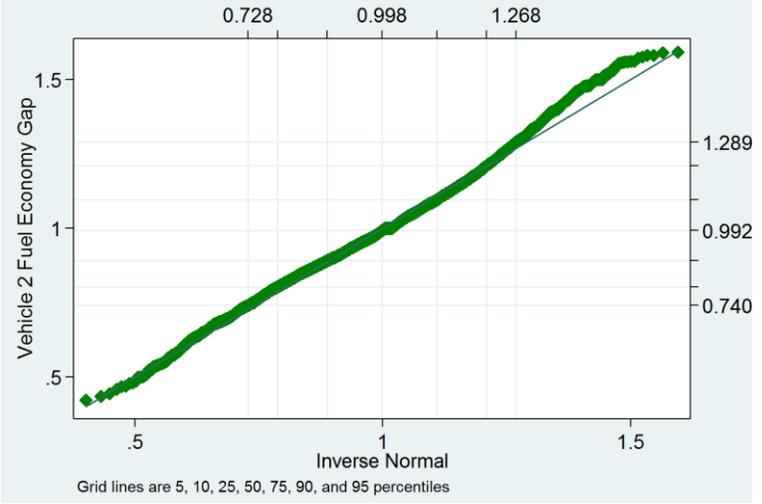

**FIGURE 4: Q-Q Plots for Fuel Economy Gaps of Vehicle 1 and 2 for the Untrimmed and Trimmed Samples**

**Notes:**
a) Y-axis values are the quantiles of the fuel economy gaps and X-axis values are the quantiles of corresponding normal distribution.
b) 5th percentile, 50th percentile, and 95th percentile values of fuel economy gaps are labelled on the second secondary y-axis in each plot.
c) 5th percentile, 50th percentile, and 95th percentile values of normal distributions re labelled on the second secondary x-axis in each plot.

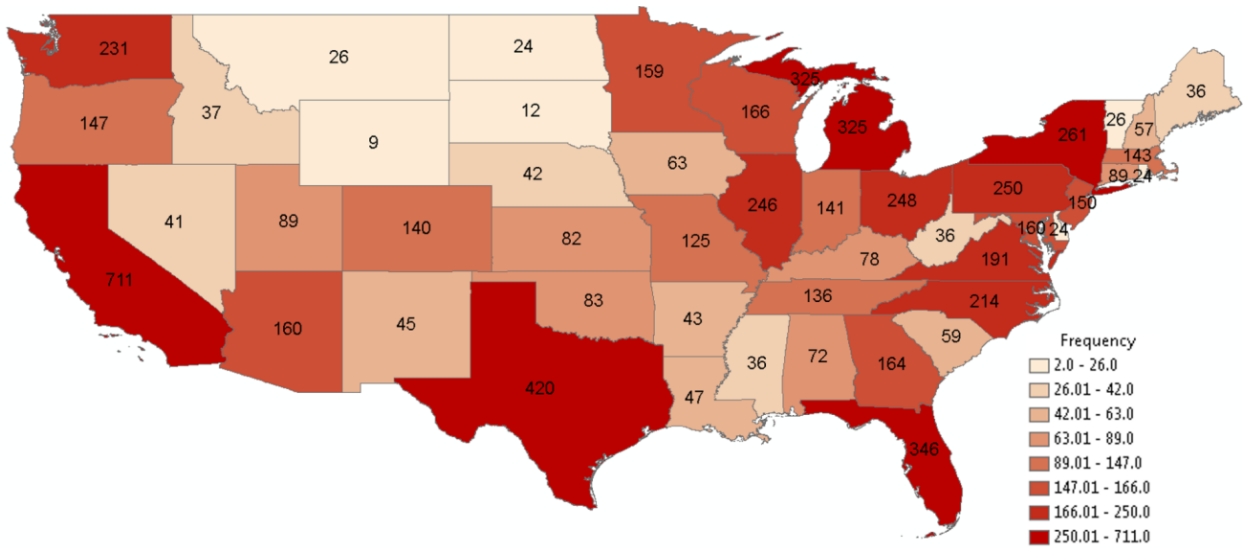

**FIGURE 5: Spatial Distribution of Sample For 6612 Garages (with data on two-vehicles) Across the US**
(Note: State information is missing for 514 garages)

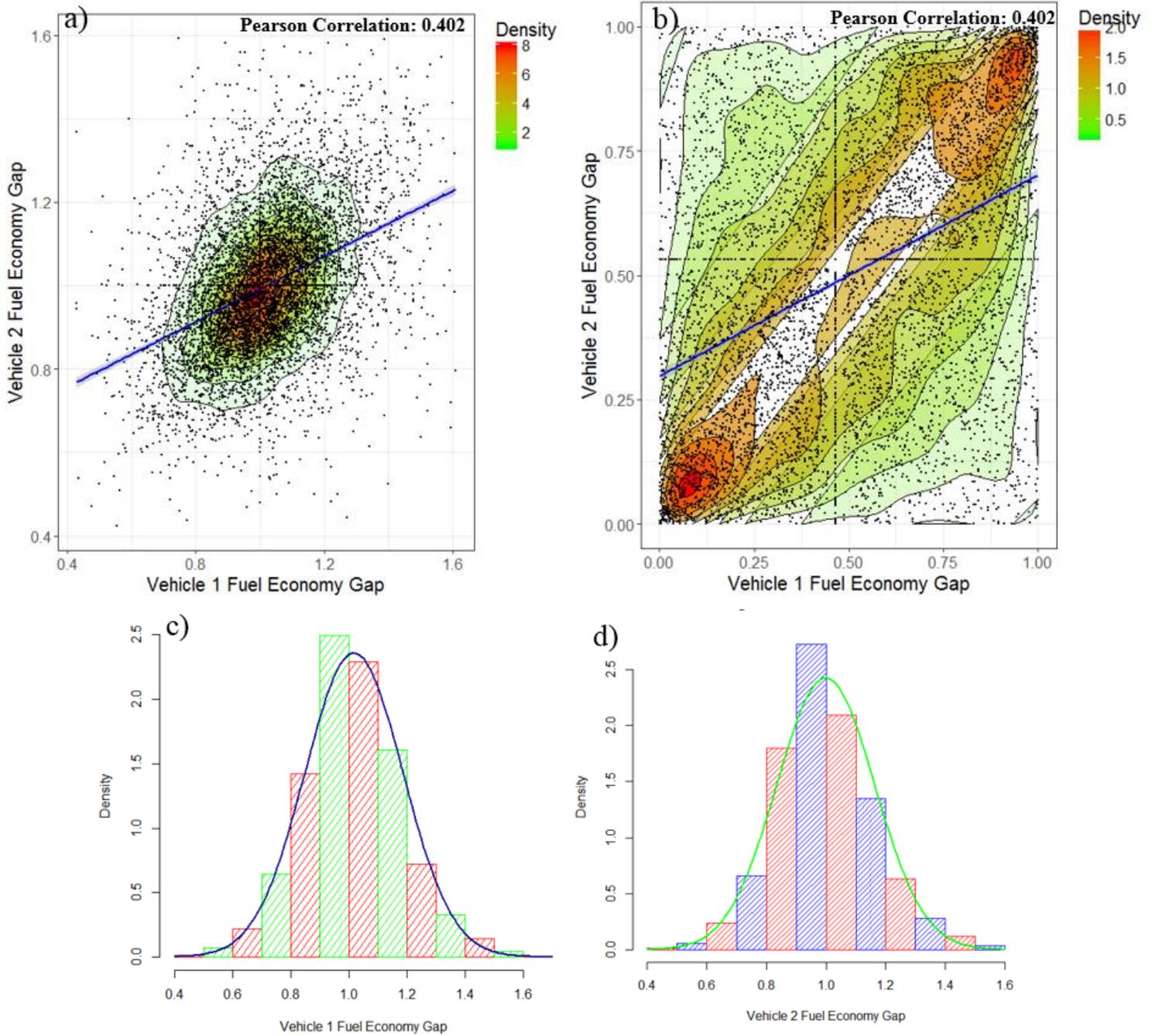

**FIGURE 6: Bivariate Distributions and Marginal Distributions of the Fuel Economy Gaps Within a Garage**

(Notes: a) Bivariate relationship between fuel economy gaps for vehicle 1 and vehicle 2; b) Bivariate relationship of transformed copula [0,1] data; c): marginal distribution of vehicle 1 fuel economy gaps; d) marginal distribution of vehicle 2 fuel economy gaps.)

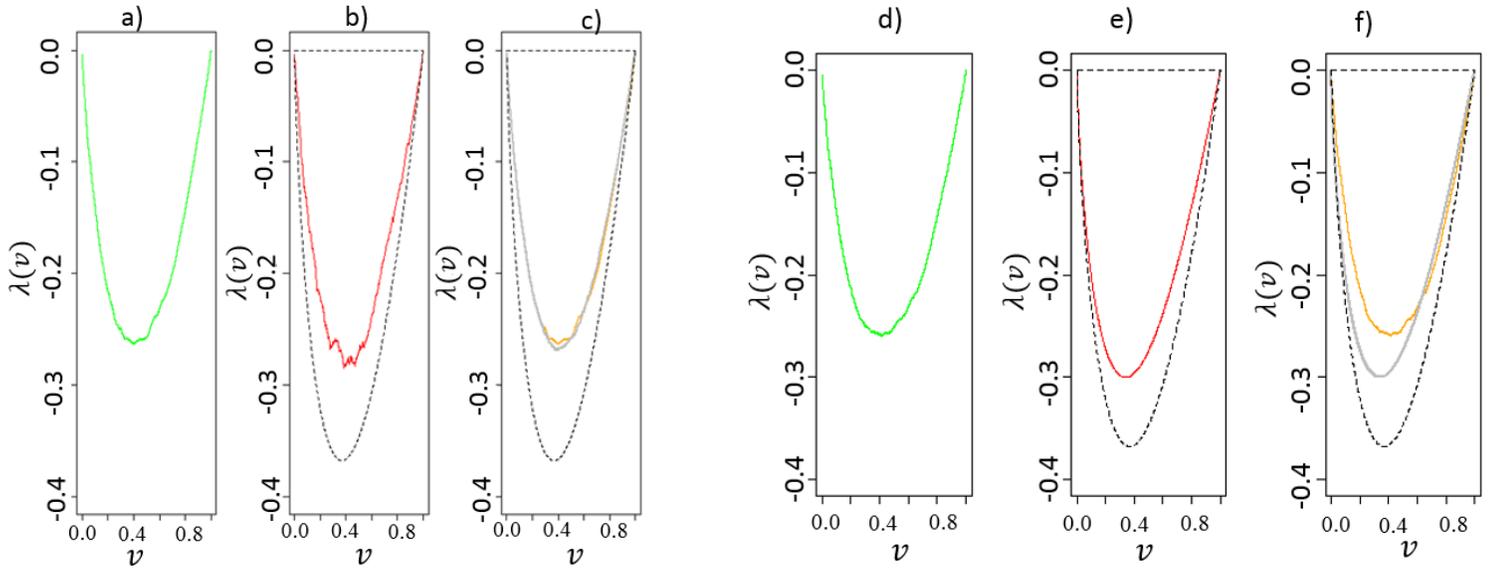

**FIGURE 7: $\lambda$-function Goodness of Fit Plots for Best-Fit Student-t and Worst Fit Joe Copula.**

**Notes:**

- a) and d) show the empirical $\lambda$-function for the bivariate fuel economy data
- b) Red series shows the theoretical $\lambda$-function for best-fit Student t-copula (with $\hat{\theta}$ of 0.427 and DF of 5.325)
- c) shows the empirical $\lambda$-function of the bivariate data (orange series) superimposed on theoretical $\lambda$-function of best-fit Student-t copula (solid grey series)
- e) shows the theoretical $\lambda$-function for worst-fit Joe copula (with $\hat{\theta}$ of 1.436)
- f) shows the empirical $\lambda$-function of the bivariate data (orange series) superimposed on theoretical $\lambda$-function of the worst-fit Joe copula (solid grey series).
- For each of (b), (c), (e), and (f), the two dotted grey series show the boundaries of the $\lambda$-function for complete independence (Kendall's $\tau = 0$) and comonotonicity (Kendall's $\tau = 1$ or $\lambda = 0$).

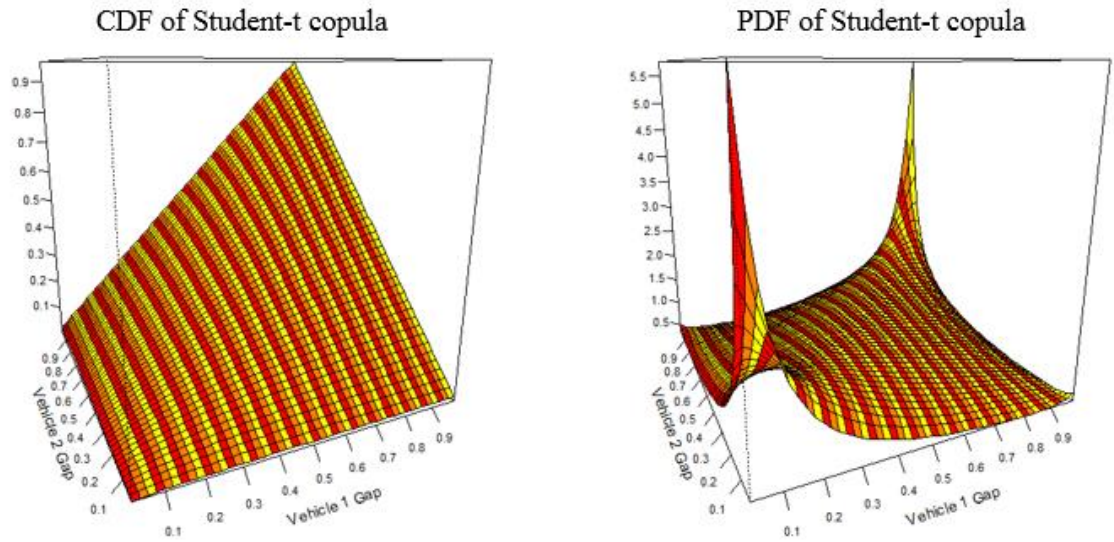

**FIGURE 8: Cumulative Distribution Function (CDF) and Probability Density Function (PDF) of Student-t Copula ($\hat{\theta}$ of 0.427 and DF of 5.325).**

(Note: Lower and upper tail dependence coefficients are 0.1594 each)

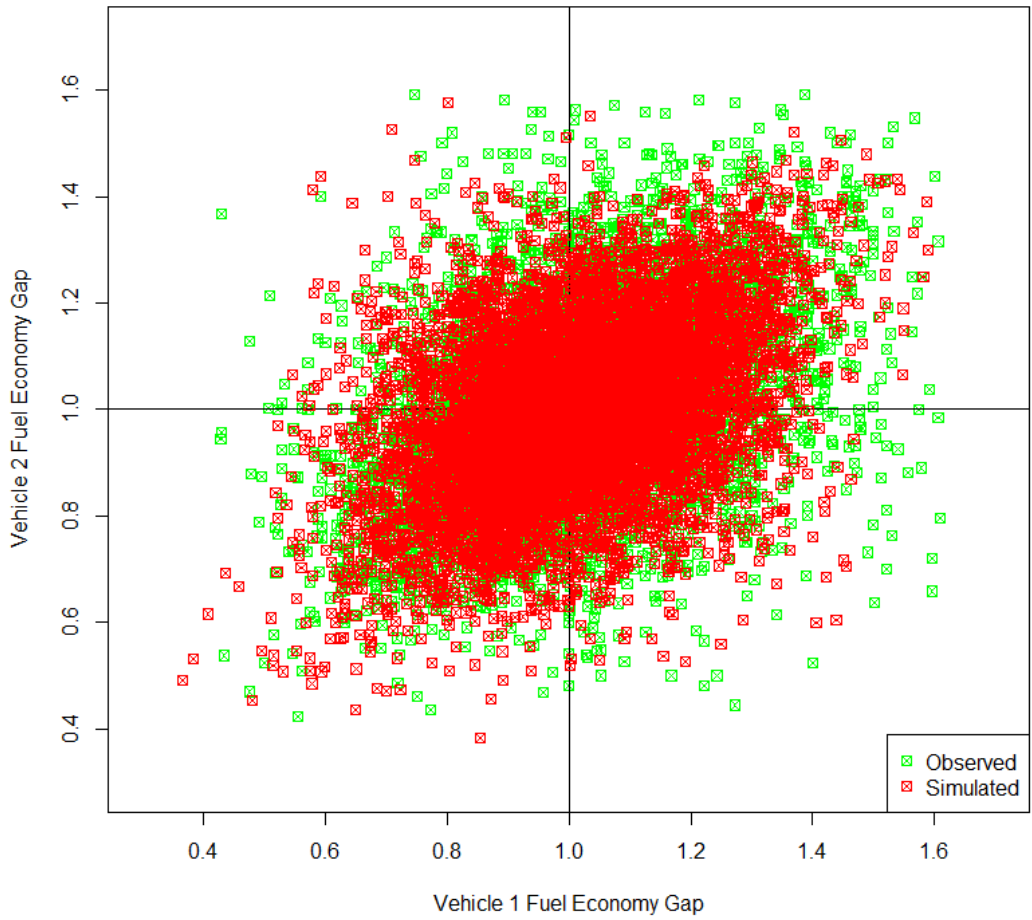

**FIGURE 9: Observed vs. Simulated (using best-fit student-copula) Fuel Economy Gaps for Vehicle 1 and 2.**

**TABLE 1** Mathematical Formulations and Characteristics of Elliptical, Archimedean, and Survival Archimedean Copulas

| | Bivariate Elliptical Copula Families | | | |
|---|---|---|---|---|
| Copula Name | Formulation | Range of parameter(s) | Kendall's $\tau$ | Tail dependence (lower, upper) |
| Gaussian | Eq. 3 | $p \in (-1,1)$ | $\frac{2}{\pi} arcsin(p)$ | 0 |
| Student-t | Eq. 4 | $p \in (-1,1), v > 2$ | $\frac{2}{\pi} arcsin(p)$ | $2t_{v+1}(-\sqrt{v+1}\sqrt{\frac{1-p}{1+p}})$ |
| | One-Parameter Archimedean Copula Families | | | |
| Copula Name | Generator Function | Range of parameter(s) | Kendall's $\tau$ | Tail dependence (lower, upper) |
| Clayton | $\frac{1}{\theta}(t^{-\theta}-1)$ | $\theta > 0$ | $\frac{\theta}{\theta+2}$ | $(2^{-\theta}, 0)$ |
| Frank | $-\log\left[\frac{e^{-\theta t}-1}{e^{-\theta}-1}\right]$ | $-\infty \leq \theta < \infty$ | $1 - \frac{4}{\theta}\{1 - D_1(\theta)\}$ | $(0,0)$ |
| Gumbel | $(-\log t)^\theta$ | $\theta \geq 1$ | $\frac{\theta-1}{\theta}$ | $(0, 2-2^{\frac{1}{\theta}})$ |
| Joe | $-\log[1-(1-t)^\theta]$ | $\theta > 1$ | $1 + \frac{4}{\theta}D_2(\theta)$ | $(0, 2-2^{\frac{1}{\theta}})$ |
| | Two-Parameter Hybrid Archimedean Copula Families | | | |
| BB1 | $(t^{-\theta}-1)^\delta$ | $\theta > 0, \delta \geq 1$ | $1 - \frac{2}{\delta(\theta+2)}$ | $(2^{-\frac{1}{\theta\delta}}, 2-2^{\frac{1}{\delta}})$ |
| BB6 | $(-\log[1-(1-t)^\theta])^\delta$ | $\theta \geq 1, \delta \geq 1$ | $1 + \frac{4}{\theta\delta}D_3(\theta)$ | $(0, 2-2^{\frac{1}{\delta\theta}})$ |
| BB7 | $(1-(1-t)^\theta)^{-\delta}-1$ | $\theta \geq 1, \delta > 0$ | $1 + \frac{4}{\theta\delta}D_4(\theta)$ | $(2^{-\frac{1}{\delta}}, 2-2^{\frac{1}{\theta}})$ |
| BB8 | $-\log\left[\frac{1-(1-\delta t)^\theta}{1-(1-\delta)^\theta}\right]$ | $\theta \geq 1, \delta \in (0,1)$ | $1 + \frac{4}{\theta\delta}D_5(\theta)$ | $(0,0)$ |

Notes:

- $D_1(\theta)$ for Frank copula is a Debye Function as: $D_1(\theta) = \frac{1}{\theta}\int_0^\theta \frac{t}{e^t-1}dt$

- $D_2(\theta)$ for Joe copula is: $D_2(\theta) = \int_{t=0}^1 t\log(t)(1-t)^{2(1-\theta)/\theta}dt$

- $D_3(\theta)$ for BB6 copula is: $D_3(\theta) = \int_{t=0}^1 (-\log(1-(1-t)^\theta) \times (1-t)(1-(1-t)^{-\theta}))dt$

- $D_4(\theta)$ for BB7 copula is: $D_4(\theta) = \int_{t=0}^1 \left(-(1-(1-t)^\theta)^{\delta+1} \times \frac{(1-(1-t)^\theta)^{-\delta}-1}{(1-t)^{\theta-1}}\right)dt$

- $D_5(\theta)$ for BB8 copula is: $D_5(\theta) = \int_{t=0}^1 \left(-\log\left(\frac{(1-t\delta)^\theta-1}{(1-\delta)^\theta-1}\right) \times (1-t\delta)(1-(1-t\delta)^{-\theta})\right)dt$

**TABLE 2: Descriptive Statistics of Key Variables**

| | Vehicle 1 | | | | Vehicle 2 | | | |
|---|---|---|---|---|---|---|---|---|
| **Variable** | **Mean** | **SD** | **Min** | **Max** | **Mean** | **SD** | **Min** | **Max** |
| ***Untrimmed Sample (N = 7244)*** | | | | | | | | |
| Average My MPG | 24.72 | 8.54 | 2.86 | 96.09 | 26.21 | 9.05 | 3 | 100.11 |
| EPA ratings | 24.26 | 7.50 | 10 | 65 | 26.38 | 8.97 | 11 | 65 |
| Fuel economy gap (%) | 1.02 | 0.20 | 0.14 | 4.18 | 1.00 | 0.20 | 0.18 | 4.58 |
| ***Trimmed Sample (N = 7126)*** | | | | | | | | |
| Average My MPG | 24.64 | 8.23 | 6.62 | 83 | 26.09 | 8.76 | 8.06 | 91.30 |
| EPA ratings | 24.29 | 7.51 | 10 | 65 | 26.40 | 8.99 | 11 | 65 |
| Fuel economy gap (%) | 1.02 | 0.17 | 0.43 | 1.61 | 1.00 | 0.16 | 0.42 | 1.59 |
| ***Fuel Type/Powertrain*** | | | | | | | | |
| Diesel | 0.02 | 0.14 | 0 | 1 | 0.03 | 0.16 | 0 | 1 |
| Gasoline | 0.94 | 0.23 | 0 | 1 | 0.90 | 0.31 | 0 | 1 |
| Hybrid | 0.04 | 0.19 | 0 | 1 | 0.08 | 0.27 | 0 | 1 |
| ***Transmission Type*** | | | | | | | | |
| Manual | 0.26 | 0.44 | 0 | 1 | 0.20 | 0.40 | 0 | 1 |
| CVT | 0.04 | 0.20 | 0 | 1 | 0.10 | 0.30 | 0 | 1 |
| Automatic (Gears >= 7 speed) | 0.69 | 0.46 | 0 | 1 | 0.69 | 0.46 | 0 | 1 |
| Automatic (Gears <= 7 speed) | 0.00 | 0.06 | 0 | 1 | 0.01 | 0.10 | 0 | 1 |
| ***Model Year*** | | | | | | | | |
| 1984-1988 | 0.05 | 0.21 | 0 | 1 | 0.004 | 0.06 | 0 | 1 |
| 1989-1993 | 0.11 | 0.31 | 0 | 1 | 0.02 | 0.15 | 0 | 1 |
| 1994-1998 | 0.22 | 0.42 | 0 | 1 | 0.08 | 0.26 | 0 | 1 |
| 1999-2003 | 0.32 | 0.47 | 0 | 1 | 0.22 | 0.41 | 0 | 1 |
| 2004-2008 | 0.25 | 0.43 | 0 | 1 | 0.47 | 0.50 | 0 | 1 |
| 2009-2014 | 0.05 | 0.22 | 0 | 1 | 0.21 | 0.41 | 0 | 1 |

Notes: SD is standard deviation; N is sample size.

**TABLE 3: Parameter(s), Confidence Intervals, Kendall's Tau, Maximum Log-Likelihood, AIC, and BIC for All 21 Copula Considered**

| Copula | $\hat{\rho}$ or $\hat{\theta}$ | CI | DF or $\hat{\delta}$ | CI | $\tau$ | $LL_{max}$ | AIC | BIC |
|---|---|---|---|---|---|---|---|---|
| Gaussian | 0.406 | [0.405,0.407] | --- | --- | 0.266 | 640.16 | -1278.32 | -1271.45 |
| Student-t | 0.427 | [0.425,0.427] | 5.325 | [5.316,5.334] | 0.281 | 770.69 | -1537.38 | -1523.64 |
| Clayton | 0.566 | [0.564,0.567] | --- | --- | 0.221 | 552.91 | -1103.82 | -1096.95 |
| Gumbel | 1.352 | [1.350,1.352] | --- | --- | 0.260 | 651.94 | -1301.89 | -1295.01 |
| Frank | 2.791 | [2.788,2.793] | --- | --- | 0.289 | 662.06 | -1322.11 | -1315.24 |
| Joe | 1.436 | [1.434,1.437] | --- | --- | 0.197 | 483.89 | -965.77 | -958.90 |
| BB1 | 0.254 | [0.252,0.254] | 1.225 | [1.224,1.226] | 0.276 | 722.72 | -1441.45 | -1427.71 |
| BB6 | 1.001 | [0.998,1.003] | 1.351 | [1.349,1.352] | 0.260 | 651.66 | -1299.32 | -1285.57 |
| BB7 | 1.271 | [1.269,1.272] | 0.418 | [0.416,0.418] | 0.264 | 706.01 | -1408.02 | -1394.28 |
| BB8 | 6.000 | [5.972,6.027] | 0.403 | [0.400,0.404] | 0.284 | 656.57 | -1309.15 | -1295.41 |
| Survival Clayton | 0.544 | [0.542,0.544] | --- | --- | 0.214 | 514.23 | -1026.45 | -1019.58 |
| Survival Gumbel | 1.359 | [1.358,1.360] | --- | --- | 0.264 | 678.53 | -1355.06 | -1348.19 |
| Survival Joe | 1.459 | [1.458,1.460] | --- | --- | 0.205 | 527.71 | -1053.41 | -1046.54 |
| Survival BB1 | 0.200 | [0.199,0.201] | 1.256 | [1.255,1.256] | 0.276 | 724.58 | -1445.16 | -1431.42 |
| Survival BB6 | 1.001 | [0.998,1.003] | 1.358 | [1.356,1.359] | 0.264 | 678.30 | -1352.59 | -1338.85 |
| Survival BB7 | 1.311 | [1.310,1.312] | 0.371 | [0.370,0.371] | 0.263 | 705.11 | -1406.21 | -1392.47 |
| Survival BB8 | 4.894 | [4.862,4.925] | 0.482 | [0.479,0.485] | 0.286 | 666.98 | -1329.96 | -1316.21 |
| Tawn Type 1 | 1.486 | [1.484,1.486] | 0.483 | [0.482,0.484] | 0.206 | 555.69 | -1107.39 | -1093.65 |
| Rotated Tawn type 1 (180 degrees) | 1.513 | [1.511,1.514] | 0.481 | [0.479,0.482] | 0.213 | 591.46 | -1178.93 | -1165.18 |
| Tawn Type 2 | 1.513 | [1.510,1.515] | 0.483 | [0.482,0.484] | 0.213 | 581.83 | -1159.65 | -1145.91 |
| Rotated Tawn type 2 (180 degrees) | 1.507 | [1.506,1.508] | 0.483 | [0.481,0.484] | 0.212 | 592.58 | -1181.15 | -1167.41 |

Notes: $\hat{\rho}$ is estimated Pearson linear-in form correlation (for Gaussian copula only); $\hat{\theta}$ is estimated dependence parameter for one-parameter copulas; DF is the degrees-of-freedom for student-t copula; $\hat{\delta}$ is the estimated second dependence parameter for two-parameter copulas; CI is 95% confidence intervals; $\tau$ is Kendall's tau; $LL_{max}$ is maximum log-likelihood; AIC – Akaike Information Criteria; BIC – Schwarz Bayesian Information Criteria.

**TABLE 4: Comparison of Detailed Summary Statistics for Observed and Simulated Fuel Economy Gaps (overall and by vehicle model years) From the Best-Fit Student-t Copula Model.**

|  | Vehicle 1 | | | Vehicle 2 | | |
|---|---|---|---|---|---|---|
| Statistic | Observed | Simulated | Absolute Difference | Observed | Simulated | Absolute Difference |
| *Overall* | | | | | | |
| Mean | 1.017 | 1.016 | 0.001 | 0.9986 | 0.9985 | 1E-04 |
| Median | 1.009 | 1.017 | 0.008 | 0.9926 | 0.997 | 0.0044 |
| 1st Quantile | 0.909 | 0.9 | 0.009 | 0.8911 | 0.8905 | 0.0006 |
| 3rd Quantile | 1.119 | 1.131 | 0.012 | 1.096 | 1.107 | 0.011 |
| Min | 0.4286 | 0.365 | 0.0636 | 0.4231 | 0.3831 | 0.04 |
| Max | 1.609 | 1.589 | 0.02 | 1.593 | 1.596 | 0.003 |
| MAD |  | 0.2332 |  |  | 0.2315 |  |
| RMSE |  | 0.2403 |  |  | 0.2331 |  |
| *Category 1\** | | | | | | |
| Mean | 1.001 | 1.01 | 0.009 | 1.007 | 0.992 | 0.015 |
| Median | 1 | 1 | 0 | 0.95 | 0.997 | 0.047 |
| 1st Quantile | 0.8629 | 0.896 | 0.0331 | 0.8818 | 0.859 | 0.0228 |
| 3rd Quantile | 1.13 | 1.121 | 0.009 | 1.149 | 1.113 | 0.036 |
| Min | 0.4757 | 0.518 | 0.0423 | 0.596 | 0.731 | 0.135 |
| Max | 1.579 | 1.52 | 0.059 | 1.444 | 1.268 | 0.176 |
| MAD |  | 0.245 |  |  | 0.204 |  |
| RMSE |  | 0.257 |  |  | 0.285 |  |
| *Category 2\** | | | | | | |
| Mean | 1.038 | 1.017 | 0.021 | 1.022 | 1.005 | 0.017 |
| Median | 1.04 | 1.017 | 0.023 | 1.028 | 1.015 | 0.013 |
| 1st Quantile | 0.9178 | 0.9061 | 0.0117 | 0.873 | 0.896 | 0.023 |
| 3rd Quantile | 1.158 | 1.137 | 0.021 | 1.153 | 1.105 | 0.048 |
| Min | 0.4783 | 0.5207 | 0.0424 | 0.526 | 0.599 | 0.073 |
| Max | 1.574 | 1.472 | 0.102 | 1.572 | 1.385 | 0.187 |
| MAD |  | 0.253 |  |  | 0.269 |  |
| RMSE |  | 0.256 |  |  | 0.239 |  |
| *Category 3\** | | | | | | |
| Mean | 1.034 | 1.017 | 0.017 | 1.047 | 1.005 | 0.042 |
| Median | 1.031 | 1.02 | 0.011 | 1.043 | 1.011 | 0.032 |
| 1st Quantile | 0.925 | 0.896 | 0.029 | 0.92 | 0.888 | 0.032 |
| 3rd Quantile | 1.143 | 1.127 | 0.016 | 1.158 | 1.109 | 0.049 |
| Min | 0.428 | 0.457 | 0.029 | 0.423 | 0.491 | 0.068 |
| Max | 1.606 | 1.588 | 0.018 | 1.581 | 1.469 | 0.112 |
| MAD |  | 0.24 |  |  | 0.262 |  |
| RMSE |  | 0.244 |  |  | 0.248 |  |

**Notes:** Category 1 is for vehicle model years between 1984-1988; Category 2 for vehicle model years between 1989-1993; Category 3 for vehicle model years between 1994-1998; MAD is mean absolute deviance; and RMSE is root mean squared error.

**TABLE 4 *(Continued)*: Comparison of Detailed Summary Statistics for Observed and Simulated Fuel Economy Gaps (overall and by vehicle model years) From the Best-Fit Student-t Copula Model.**

|  | Vehicle 1 | | | Vehicle 2 | | |
|---|---|---|---|---|---|---|
| **Statistic** | **Observed** | **Simulated** | **Absolute Difference** | **Observed** | **Simulated** | **Absolute Difference** |
| *Category 4* | | | | | | |
| Mean | 1.023 | 1.016 | 0.007 | 1.03 | 0.999 | 0.031 |
| Median | 1.019 | 1.02 | 0.001 | 1.018 | 1.004 | 0.014 |
| 1st Quantile | 0.925 | 0.901 | 0.024 | 0.919 | 0.89 | 0.029 |
| 3rd Quantile | 1.121 | 1.127 | 0.006 | 1.133 | 1.108 | 0.025 |
| Min | 0.428 | 0.365 | 0.063 | 0.444 | 0.474 | 0.03 |
| Max | 1.609 | 1.589 | 0.02 | 1.582 | 1.521 | 0.061 |
| MAD | | 0.228 | | | 0.24 | |
| RMSE | | 0.236 | | | 0.233 | |
| *Category 5* | | | | | | |
| Mean | 0.9981 | 1.017 | 0.0189 | 0.998 | 0.998 | 0 |
| Median | 0.996 | 1.016 | 0.02 | 0.992 | 0.996 | 0.004 |
| 1st Quantile | 0.894 | 0.9 | 0.006 | 0.895 | 0.893 | 0.002 |
| 3rd Quantile | 1.09 | 1.136 | 0.046 | 1.091 | 1.106 | 0.015 |
| Min | 0.506 | 0.384 | 0.122 | 0.436 | 0.435 | 0.001 |
| Max | 1.606 | 1.519 | 0.087 | 1.591 | 1.576 | 0.015 |
| MAD | | 0.229 | | | 0.225 | |
| RMSE | | 0.233 | | | 0.23 | |
| *Category 6* | | | | | | |
| Mean | 0.9547 | 1.007 | 0.0523 | 0.947 | 0.995 | 0.048 |
| Median | 0.94 | 0.998 | 0.058 | 0.944 | 0.986 | 0.042 |
| 1st Quantile | 0.856 | 0.889 | 0.033 | 0.849 | 0.888 | 0.039 |
| 3rd Quantile | 1.052 | 1.124 | 0.072 | 1.037 | 1.109 | 0.072 |
| Min | 0.562 | 0.436 | 0.126 | 0.46 | 0.383 | 0.077 |
| Max | 1.444 | 1.463 | 0.019 | 1.593 | 1.551 | 0.042 |
| MAD | | 0.223 | | | 0.224 | |
| RMSE | | 0.227 | | | 0.23 | |

**Notes:** Category 4 is for vehicle model years between 1999-2003; Category 5 for vehicle model years between 2004-2008; Category 6 for vehicle model years between 2009-2014; MAD is mean absolute deviance; and RMSE is root mean squared error.

**TABLE 5: Conditional Distributions of the Fuel Economy Gaps of Vehicle 1 and 2 Based on Observed Bivariate Data**

| Vehicle 1 \ Vehicle 2 | At least 10% better than label values[a] | At most 10% better than label values[b] | At most 10% worse than label values[c] | At least 10% worse than label values[d] | Total |
|---|---|---|---|---|---|
| **At least 10% better than label values[a]** | 881 | 483 | 451 | 271 | 2,086 |
|  | 42.23 | 23.15 | 21.62 | 12.99 | 100 |
|  | 50.37 | 33.06 | 23.21 | 13.74 | 29 |
| **At most 10% better than label values[b]** | 403 | 410 | 450 | 313 | 1,576 |
|  | 25.57 | 26.02 | 28.55 | 19.86 | 100 |
|  | 23.04 | 28.06 | 23.16 | 15.86 | 22 |
| **At most 10% worse than label values[c]** | 262 | 361 | 617 | 539 | 1,779 |
|  | 14.73 | 20.29 | 34.68 | 30.3 | 100 |
|  | 14.98 | 24.71 | 31.76 | 27.32 | 25 |
| **At least 10% worse than label values[d]** | 203 | 207 | 425 | 850 | 1,685 |
|  | 12.05 | 12.28 | 25.22 | 50.45 | 100 |
|  | 11.61 | 14.17 | 21.87 | 43.08 | 24 |
| **Total** | 1,749 | 1,461 | 1,943 | 1,973 | 7,126 |
|  | 24.54 | 20.5 | 27.27 | 27.69 | 100 |
|  | 100 | 100 | 100 | 100 | 100 |

Notes: The number at the top of each cell is the frequency count (shown in white cells); the second number is the row percentage—they sum to 100% going across the table (indicated by light grey cells), i.e., for a given fuel economy gap category of vehicle 1, it shows the distribution of different categories of fuel economy gaps for vehicle 2; and the third number is the column percentage—they sum to 100% going down the table (indicated by dark grey cells), i.e., for a given fuel economy gap category of vehicle 2, it shows the distribution of different categories of fuel economy gaps for vehicle 1; (a) indicates fuel economy gaps of greater than 1.1; (b) indicates fuel economy gaps between 1 and 1.1; (c) indicates fuel economy gaps between 0.9 and 1; and (d) fuel economy gaps of less than or equal to 0.9.

TABLE 6: Conditional Distributions of the Fuel Economy Gaps of Vehicle 1 and 2 Based on Predicted Bivariate Fuel Economy Data from the Best-Fit Student-t Copula Model.

| Vehicle 1 \ Vehicle 2 | At least 10% better than label values[a] | At most 10% better than label values[b] | At most 10% worse than label values[c] | At least 10% worse than label values[d] | Total |
|---|---|---|---|---|---|
| **At least 10% better than label values[a]** | 979 | 518 | 415 | 286 | 2,198 |
|  | 44.54 | 23.57 | 18.88 | 13.01 | 100 |
|  | 51.61 | 32.19 | 24.53 | 14.83 | 30.84 |
| **At most 10% better than label values[b]** | 400 | 476 | 448 | 334 | 1,658 |
|  | 24.13 | 28.71 | 27.02 | 20.14 | 100 |
|  | 21.09 | 29.58 | 26.48 | 17.32 | 23.27 |
| **At most 10% worse than label values[c]** | 272 | 352 | 419 | 445 | 1,488 |
|  | 18.28 | 23.66 | 28.16 | 29.91 | 100 |
|  | 14.34 | 21.88 | 24.76 | 23.08 | 20.88 |
| **At least 10% worse than label values[d]** | 246 | 263 | 410 | 863 | 1,782 |
|  | 13.8 | 14.76 | 23.01 | 48.43 | 100 |
|  | 12.97 | 16.35 | 24.23 | 44.76 | 25.01 |
| **Total** | 1,897 | 1,609 | 1,692 | 1,928 | 7,126 |
|  | 26.62 | 22.58 | 23.74 | 27.06 | 100 |
|  | 100 | 100 | 100 | 100 | 100 |

Notes: The number at the top of each cell is the frequency count (shown in white cells); the second number is the row percentage—they sum to 100% going across the table (indicated by light grey cells), i.e., for a given fuel economy gap category of vehicle 1, it shows the distribution of different categories of fuel economy gaps for vehicle 2; and the third number is the column percentage—they sum to 100% going down the table (indicated by dark grey cells), i.e., for a given fuel economy gap category of vehicle 2, it shows the distribution of different categories of fuel economy gaps for vehicle 1; (a) indicates fuel economy gaps of greater than 1.1; (b) indicates fuel economy gaps between 1 and 1.1; (c) indicates fuel economy gaps between 0.9 and 1; and (d) fuel economy gaps of less than or equal to 0.9.